\documentclass[%
 preprint,
showpacs,
nofootinbib,
 amsmath,amssymb,
 prc,
 superscriptaddress
]{revtex4-1}

\usepackage{graphicx}% Include figure files
\usepackage{dcolumn}% Align table columns on decimal point
\usepackage{bm}% bold math
\usepackage{braket}
\usepackage{caption}
\usepackage{subcaption}
\begin{document}

%\preprint{APS/123-QED}

\title{Cluster-Daughter Overlap as a New Probe of Alpha-Cluster Formation in Medium-Mass and Heavy Even-Even Nuclei}% Force line breaks with \\


\author{Dong Bai}
 \email{dbai@itp.ac.cn}
\affiliation{School of Physics, Nanjing University, Nanjing, 210093, China}%

\author{Zhongzhou Ren}
\email{Corresponding Author: zren@tongji.edu.cn}
\affiliation{School of Physics Science and Engineering, Tongji University, Shanghai, 200092, China}%
%\affiliation{School of Physics, Nanjing University, Nanjing, 210093, China}%


\date{\today}% It is always \today, today,

\begin{abstract}
We study the possibility to use the cluster-daughter overlap as a new probe of alpha-cluster formation in medium-mass and heavy even-even nuclei. We introduce a dimensionless parameter $\mathfrak{O}$, which is the ratio between the root-mean-square (rms) intercluster separation and the sum of rms point radii of the daughter nucleus and the alpha particle, to measure the degree of the cluster-daughter overlap quantitatively. By using this parameter, a large (small) cluster-daughter overlap between the alpha cluster and daughter nucleus corresponds to a small (large) $\mathfrak{O}$ value. The alpha-cluster formation is shown explicitly, in the framework of the quartetting wave function approach, to be suppressed when the $\mathfrak{O}$ parameter is small, and be favored when the $\mathfrak{O}$ parameter is large. We then use this $\mathfrak{O}$ parameter to explore systematically the landscape of alpha-cluster formation probabilities in medium-mass and heavy even-even nuclei, with $\mathfrak{O}$ being calculated from experimentally measured charge radii. The trends of alpha-cluster formation probabilities are found to be generally consistent with previous studies. The effects of various shell closures on the alpha-cluster formation are identified, along with some hints on a possible subshell structure at $N=106$ along the Hg and Pb isotopic chains. The study here could be a useful complement to the traditional route to probe alpha-cluster formation in medium-mass and heavy even-even nuclei using alpha-decay data. Especially, it would be helpful in the cases where the target nucleus is stable against alpha decay or alpha-decay data are currently not available.
\end{abstract}

%\pacs{21.10.Dr, 21.10.Gv, 03.75.Hh}% PACS, the Physics and Astronomy
\maketitle

%\tableofcontents

\section{INTRODUCTION}

Alpha-cluster formation plays a key role in understanding structures and reactions of light, medium-mass, and heavy/superheavy nuclei across the nuclide chart \cite{Beck:2010,Beck:2012,Beck:2014,Freer:2017gip,Delion:2010,Horiuchi:2012}. For light nuclei, the existence of alpha-cluster structures was suggested in the late 1930s \cite{Wheeler:1937zza,Hafstad:1938}. This idea is then explored in detail by generations of nuclear physicists, and inspires many important theoretical developments, such as resonating group method \cite{Wheeler:1937zz}, generator coordinate method \cite{Brink:1966}, antisymmetrized molecular dynamics \cite{Kanada-Enyo:2001yji}, orthogonality condition method \cite{Saito:1969zz}, THSR (Tohsaki-Horiuchi-Schuck-R\"opke) wave function \cite{Tohsaki:2001an}, etc. It is found that there could be alpha-cluster structures in ground states of light nuclei such as $^8$Be $= \alpha + \alpha$, $^{20}$Ne $ = \alpha + {}^{16}$O, $^{44}$Ti $ = \alpha + {}^{40}$Ca \cite{Horiuchi:2012}, as well as in the famous Hoyle and Hoyle-like excited states of self-conjugate nuclei near alpha-particle disintegration thresholds \cite{Tohsaki:2001an,Funaki:2008gb,Funaki:2009fc,Yamada:2003cz,Bai:2018gqt}. The study of alpha-cluster formation in heavy/superheavy nuclei could date back to Rutherford's discovery of alpha decay more than one hundred years ago \cite{Rutherford:1899}. Till today, alpha decay is still an important direction being continuously developed \cite{Buck:1992zz,Delion:2010,Delion:2018rrl,Xu:2005ukj,Delion:2017ozx,Bai:2018adq}, from which we gain lots of information on alpha-cluster formation in heavy/superheavy nuclei \cite{Zhongzhou:1988zz,Zhongzhou:1987zz,Zhang:2008ct,Zhang:2009zzs,Qi:2014ska,Khuyagbaatar:2015hjj,Andreyev:2013iwa,Hatsukawa:1990zz,Brown:1992rg,Qian:2018dyz,Batchelder:1997zz,Xu:2006fq,Denisov:2009ng,Buck:1992zz,Delion:2004wv}. Especially, it is found that alpha-cluster formation probabilities change significantly at magic numbers $N=126$ and $Z=82$. Alpha-cluster formation could also exist in the medium-mass nuclei. Possible candidates include $^{46,54}\text{Cr}=\alpha+{}^{42,50}\text{Ti}$ \cite{Souza:2016bpt,Mohr:2017zot}, $^{90}\text{Sr}=\alpha+{}^{86}\text{Kr}$ \cite{Souza:2015yxa}, $^{92}\text{Zr}=\alpha+{}^{88}\text{Sr}$ \cite{Souza:2015yxa}, $^{94}\text{Mo}=\alpha+{}^{90}\text{Zr}$ \cite{Ohkubo:1995zz,Souza:2015yxa,Buck:1995zza,Michel:2000nw}, $^{96}\text{Ru}=\alpha+{}^{92}\text{Mo}$ \cite{Souza:2015yxa}, $^{98}\text{Pd}=\alpha+{}^{94}\text{Ru}$ \cite{Souza:2015yxa}, $^{136}\text{Te}=\alpha+{}^{132}\text{Sn}$ \cite{Wang:2013gk}, etc. Recently, there are also systematic studies on the landscape of alpha-cluster formation probabilities in medium-mass nuclei using alpha-decay data \cite{Qian:2018bfu}.

Unlike alpha clustering in light nuclei which could be studied by using microscopic methods, probing alpha clustering in medium-mass and heavy/superheavy nuclei is more challenging thanks to intrinsic difficulties in solving quantum systems with a large number of nucleons. In literature, alpha-cluster formation in medium-mass and heavy/superheavy nuclei is usually investigated systematically by exploiting experimental data on alpha decay. However, neither all the nuclei with alpha-cluster structures permit alpha decays, nor all the nuclei permitting alpha decays have their decay widths be measured. Therefore, new probes of alpha-cluster formation besides those based on the alpha decay are necessary. In this note, we would like to study the possibility to adopt the cluster-daughter overlap as a new probe of alpha-cluster formation in medium-mass and heavy nuclei, motivated by the observation that alpha-cluster formation is suppressed when the cluster-daughter overlap is large and favored when the cluster-daughter overlap is small. In Section \ref{ISvsACF}, we analyze the relation between the cluster-daughter overlap and the alpha-cluster formation quantitatively within the framework of the quartetting wave function approach \cite{Ropke:2014wsa,Xu:2017vyt,Xu:2015pvv}, which is a microscopic method proposed recently to estimate alpha-cluster formation probabilities. The inclusion of shell-model properties in the quartetting wave function approach is also discussed recently in Ref.~\cite{Ropke:2017qck}. We introduce a dimensionless parameter $\mathfrak{O}$ to measure quantitatively the degree of the cluster-daughter overlap. In Section \ref{LANDSCAPE}, the $\mathfrak{O}$ parameter is then adopted to explore the landscape of alpha-cluster formation probabilities in medium-mass and heavy even-even nuclei, with $\mathfrak{O}$ being calculated directly from experimentally measured charge radii. Special attentions are paid to the closed-(sub)shell effects. In Section \ref{CONCLUSIONS}, we give the conclusions. 

\section{CLUSTER-DAUGHTER OVERLAP VERSUS ALPHA-CLUSTER FORMATION IN THE LIGHT OF QUARTETTING WAVE FUNCTION APPROACH}
\label{ISvsACF}

\begin{figure}
\centering
\includegraphics[width=0.45\textwidth]{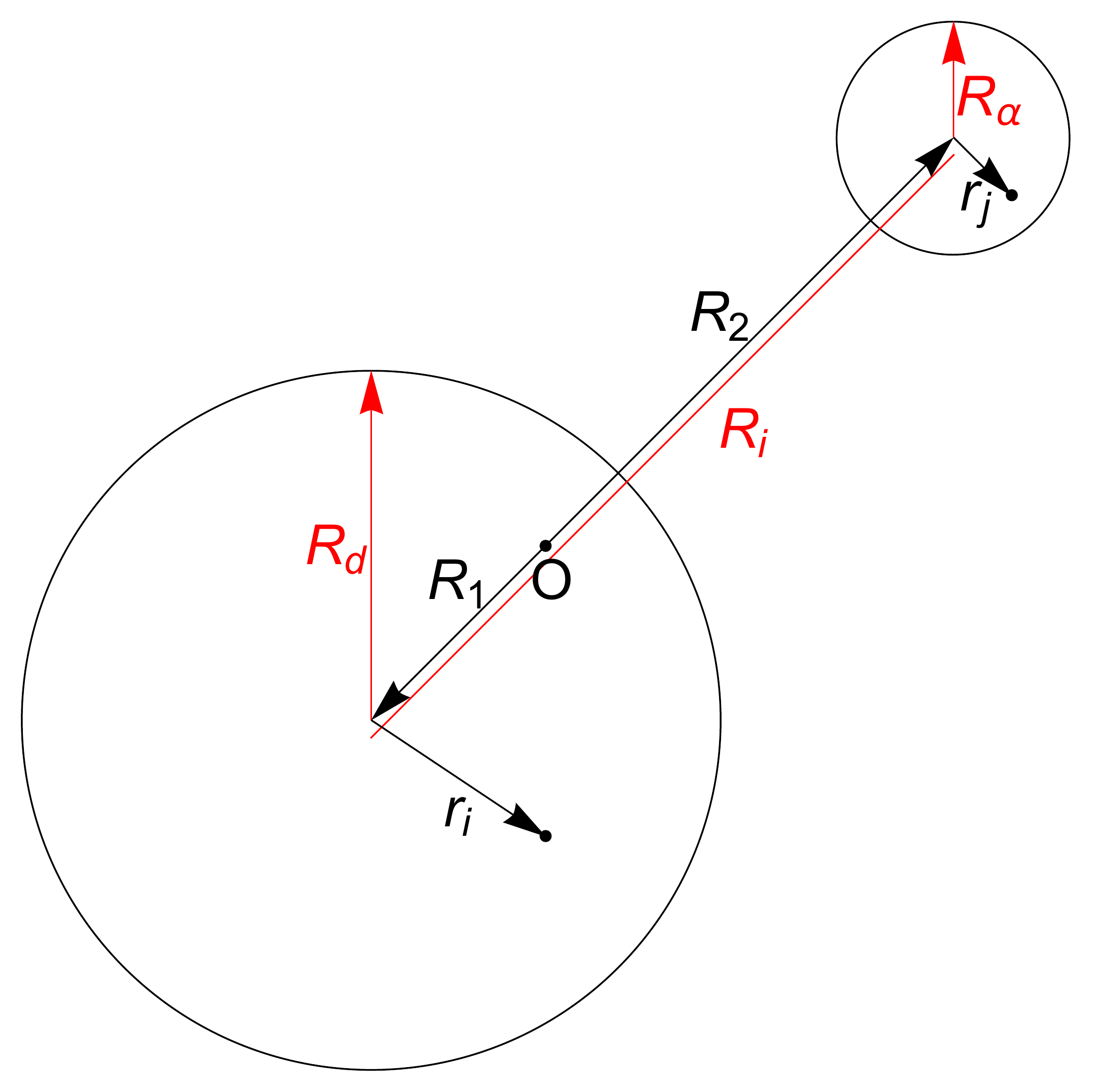}
\caption{A schematic representation of the cluster-daughter system. In this work, we assume that the centers of mass of the alpha particle and the daughter nucleus coincide with their centers of charge. $R_\alpha$ and $R_d$ are the rms point radii of the alpha particle and the daughter nucleus. $R_i$ is the rms intercluster separation. $O$ is the center of mass of the parent nucleus. $\mathbf{R}_1$ ($\mathbf{R}_2$) is the displacement vector between $O$ and the center of mass of the daughter nucleus (alpha particle). $\mathbf{r}_i$ ($\mathbf{r}_j$) is the displacement vector between the center of mass of the daughter nucleus (alpha particle) to a representative nucleon inside the daughter nucleus (alpha particle).}
\label{Demo}
\end{figure}

We first study the relation between the cluster-daughter overlap and the alpha-cluster formation probability within the framework of the quartetting wave function approach. The quartetting wave function approach is inspired by the THSR wave function for light self-conjugate nuclei, and has been applied successfully in studying alpha-cluster formation in heavy/superheavy nuclei. Within this framework, we divide the wave function of the four valence nucleons uniquely as the product of the center-of-mass (COM) part and the intrinsic part. Following Ref.~\cite{Ropke:2014wsa,Xu:2017vyt,Xu:2015pvv,Ropke:2017qck}, by adopting the local-density approximation and ignoring the derivative terms of the intrinsic wave function, the Schr\"odinger equation for the COM wave function could be given by
\begin{align}
&-\frac{\hbar^2}{2M_\alpha}\nabla^2_\mathbf{r}\Phi(\mathbf{r})
%-\frac{\hbar^2}{\mu_\alpha}\int\mathrm{d}\mathbf{s}_j\,\varphi^{\text{intr},*}(\mathbf{s}_j,\mathbf{R})[\nabla_\mathbf{R}\varphi^\text{intr}(\mathbf{s}_j,\mathbf{R})][\nabla_\mathbf{R}\Phi(\mathbf{R})]
%
%-\frac{\hbar^2}{2\mu_\alpha}\int\mathrm{d}\mathbf{s}_j\,\varphi^{\text{intr},*}(\mathbf{s}_j,\mathbf{R})[\nabla^2_\mathbf{R}\varphi^\text{intr}(\mathbf{s}_j,\mathbf{R})]\Phi(\mathbf{R})
+W(\mathbf{r})\Phi(\mathbf{r})=E\Phi(\mathbf{r}).\label{SE4COM}
%&W(\mathbf{r})=\int\mathrm{d}\mathbf{s}_j\,\varphi^{\text{intr},*}(\mathbf{s}_j,\mathbf{r})\left[T[\nabla_{\mathbf{s}_j}]+V(\mathbf{r},\mathbf{s}_j)\right]\varphi^{\text{intr}}(\mathbf{s}_j,\mathbf{r}),\label{Def4W}
\end{align}
$W(\mathbf{r})$ is the effective potential felt by the COM motion of the alpha cluster.  A detailed derivation of Eq.~\eqref{SE4COM} could be found in Ref.~\cite{Ropke:2014wsa}. The alpha-cluster formation probability $P_\alpha$ could be then obtained as follows:
\begin{align}
P_\alpha=\int\mathrm{d}^3\mathbf{r}\, |\Phi(\mathbf{r})|^2\,\Theta[\rho_B^\text{Mott}-\rho_B({\mathbf{r}})].
\label{ACFP}
\end{align}
Here, $\rho_B^\text{Mott}=0.02917\text{ fm}^{-3}$ is the Mott density higher than which the the alpha cluster is believed to dissolve and merge with the daughter nucleus. The root-mean-square (rms) intercluster separation is given by
\begin{align}
R_i\equiv \braket{R^2}^{1/2}_i=\left[\int_0^\infty\mathrm{d}r\,r^2\,\left|\phi(r)\right|^2\right]^{1/2},
\end{align}
where $\phi(r)$ is the normalized radial wave function of $\Phi(\mathbf{r})$. To measure the degree of the cluster-daughter overlap quantitatively, we introduce a dimensionless parameter $\mathfrak{O}$ to measure the relative size between the rms intercluster separation $R_i$ and the sum of the rms point radii of the alpha particle $R_\alpha\equiv\braket{R^2}_\alpha^{1/2}$ and the daughter nucleus $R_d\equiv\braket{R^2}_d^{1/2}$ (see Fig.~\ref{Demo} for a schematic representation),
\begin{align}
\mathfrak{O}=\frac{R_i}{R_\alpha+R_d}.
\label{DDef}
\end{align}
As a result, a large (small) overlap between the alpha cluster and the daughter nucleus corresponds to a small (large) value of the $\mathfrak{O}$ parameter.

In the rest part of this section, we would like to adopt the quartetting wave function approach to study the relation between the $\mathfrak{O}$ parameter and the alpha-cluster formation probability. We take $^{212}\text{Po}={}^{208}\text{Pb}+\alpha$ as an example, which could be viewed as an alpha cluster moving on the top of the doubly magic nucleus $^{208}$Pb and has been investigated intensively by previous studies \cite{Ohkubo:1995zz,Astier:2009bs,Ibrahim:2010zz,Buck:1995zza,Buck:1996zza,Varga:1992zz}. For the density distributions for the daughter nucleus, we adopt \cite{Tarbert:2013jze,Xu:2015pvv}
\begin{align}
&\rho_\pi(r)=\frac{0.0628948}{1+\exp[(r-6.68 \text{ fm})/0.447 \text{ fm}]}\text{ fm}^{-3},\\
&\rho_\nu(r)=\frac{0.0937763}{1+\exp[(r-6.70 \text{ fm})/0.550 \text{ fm}]}\text{ fm}^{-3}.
\end{align}
as the proton and neutron distributions, respectively. The critical radius that marks the dissolution of the alpha cluster is then given by $r_\text{cluster}=7.43825$ fm. In other words, the alpha cluster persists only for $r>r_\text{cluster}$ and dissolves once $r<r_\text{cluster}$. When $r>r_\text{cluster}$, the effective potential $W(\mathbf{r})=W^\text{ext}(\mathbf{r})+W^\text{intr}(\mathbf{r})$, with $W^\text{ext}(\mathbf{r})$ being the external potential inside which the alpha cluster moves and $W^\text{intr}(\mathbf{r})$ being the intrinsic energy of the alpha cluster inside the nuclear medium. The external potential $W^\text{ext}(\mathbf{r})$ could be approximated by $W^\text{ext}(\mathbf{r})=V_\text{M3Y}(\mathbf{r})+V_\text{C}(\mathbf{r})$ following the double-folding procedure. The microscopic M3Y nucleon-nucleon interaction takes the form $V_\text{NN}(s)=c\exp(-4s)/(4s)-d\exp(-2.5s)/(2.5s)$. For the alpha emitter $^{212}$Po, the two free parameters $c$ and $d$ could be determined by fitting the emission energy $Q_\alpha$ and the alpha-decay half-life $T_{1/2}$. The intrinsic potential $W^\text{intr}(\mathbf{r})$, on the other hand, is given by $W^\text{intr}(\mathbf{r})=W^\text{Pauli}(\mathbf{r})+E_\alpha^{(0)}$, with $W^\text{Pauli}(\mathbf{r})=4515.9\rho_B(\mathbf{r})-100935\rho^2_B(\mathbf{r})+1202538\rho^3_B(\mathbf{r})$ \cite{Ropke:2014wsa} and $E_\alpha^{(0)}=-28.3$ MeV being the energy of the alpha particle in the vacuum. When $r<r_\text{cluster}$, the cluster state of the four valence nucleons merges with the shell-model state of the daughter nucleus. In the local density approximation, we have $W(\mathbf{r})=\mu_4$, with $\mu_4$ being a constant related to the Fermi energy of the four valence nucleons. 

\begin{figure}

\centering

\begin{subfigure}[b]{10.5cm}
\centering
\includegraphics[width=10.5cm]{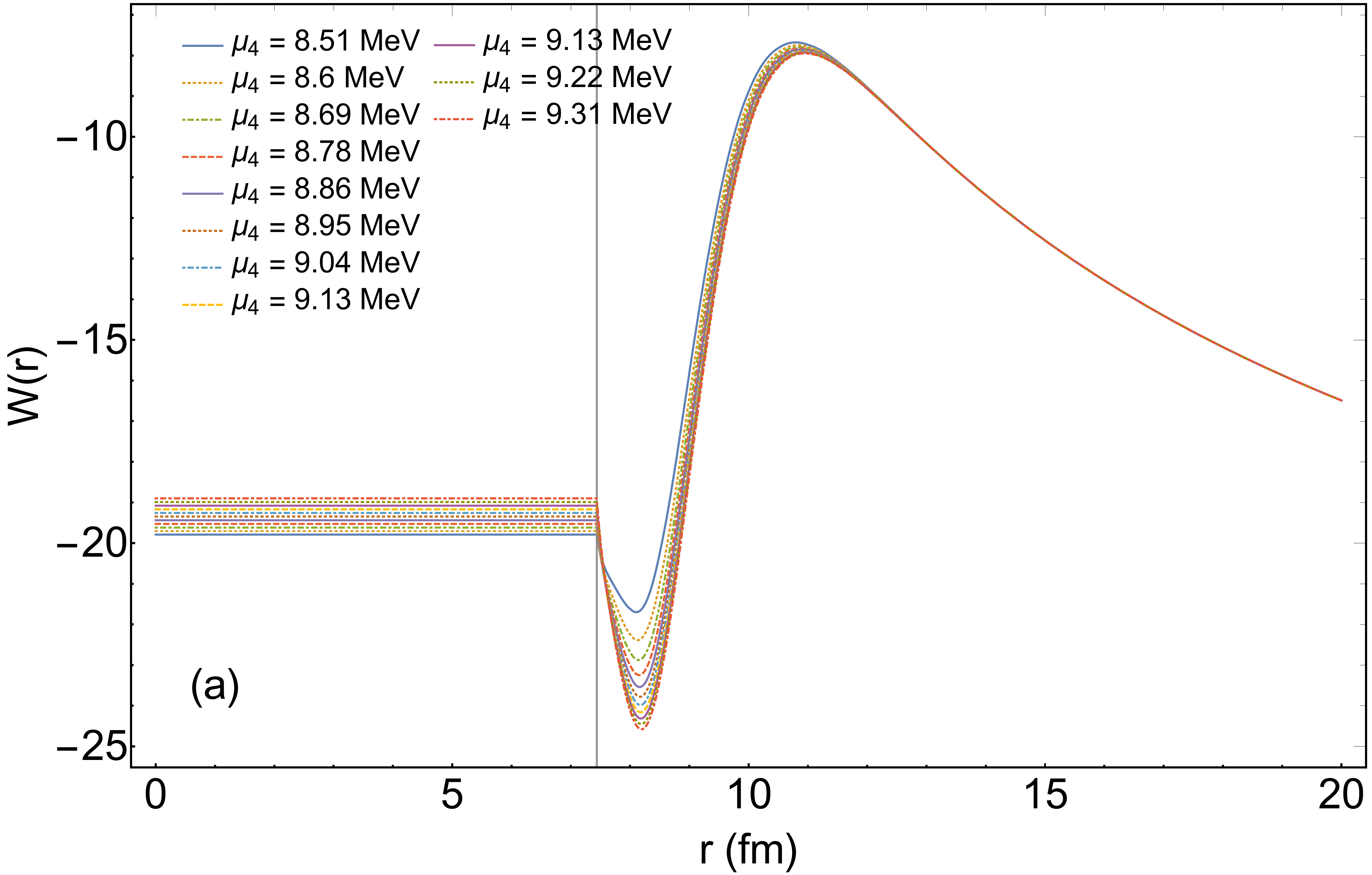}
%\caption*{$\quad\quad\quad$(a)}
\end{subfigure}

\begin{subfigure}[b]{10.5cm}
\centering
\includegraphics[width=10.5cm]{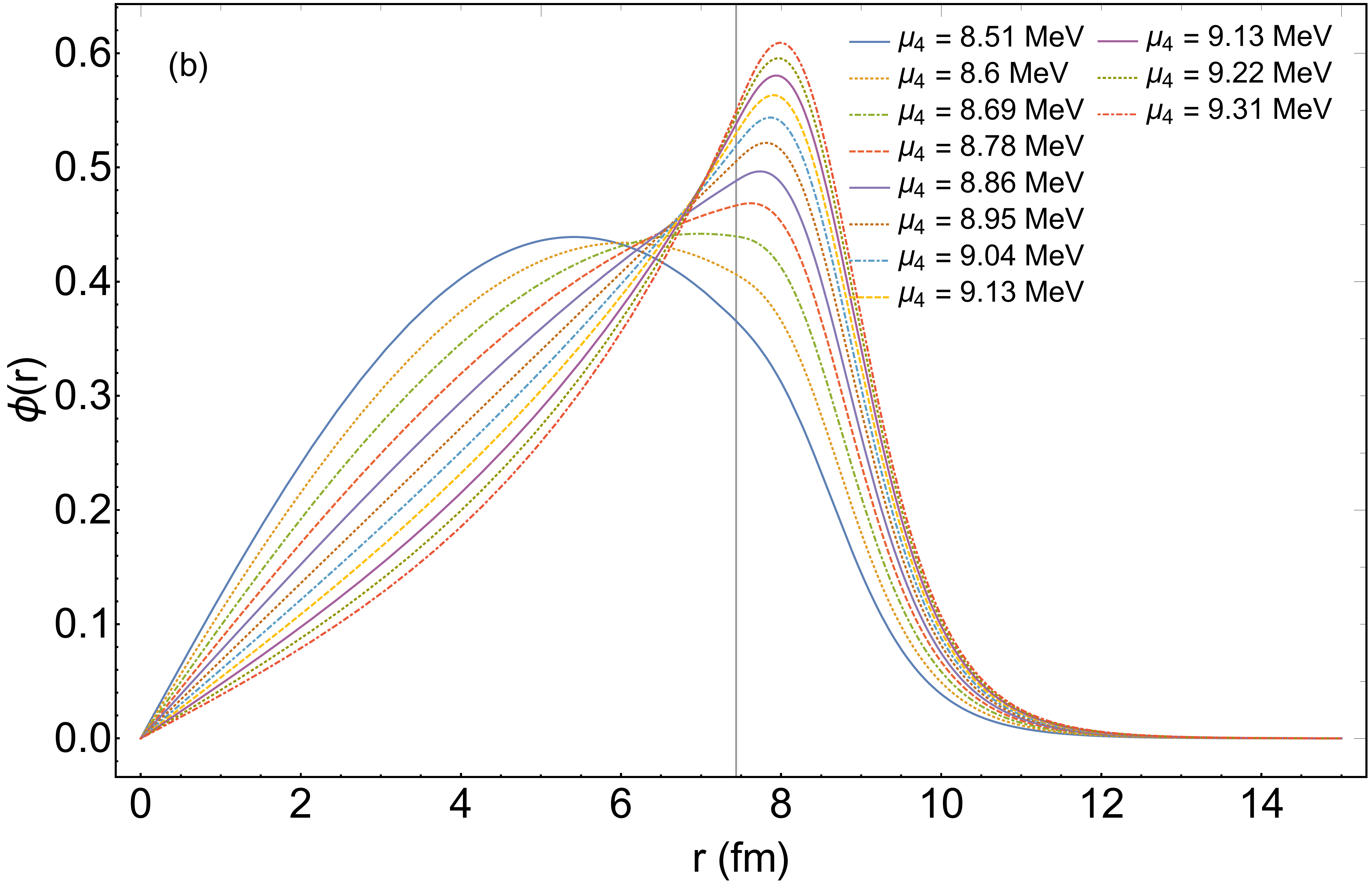}
%\caption*{$\quad\quad\quad$(b)}
\end{subfigure}

\begin{subfigure}[b]{10.5cm}
\centering
\includegraphics[width=10.5cm]{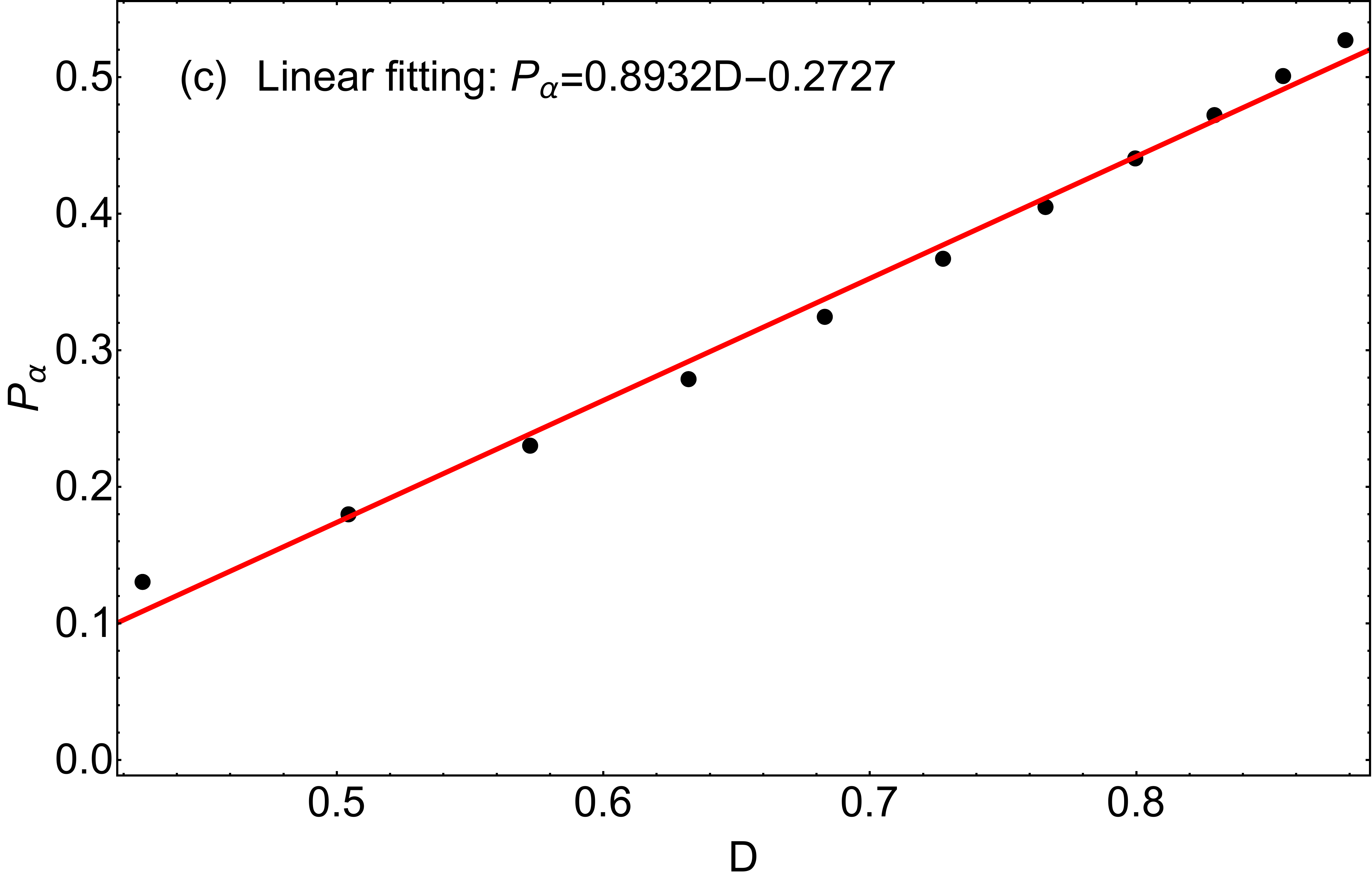}
%\caption*{$\quad\quad\quad$(c)}
\end{subfigure}

\caption{A study on $^{212}$Po within the quartetting wave function approach by varying values of $\mu_4$. Fig.~(1a) shows the behavior of the effective potential $W(r)$. Fig.~(1b) shows the normalized radial wave function $\phi(r)$. Fig.~(1c) shows the relation between the dimensionless parameter $\mathfrak{O}$ and the alpha-cluster formation probability $P_\alpha$. The critical radius $r_\text{cluster}=7.43825\text{ fm}$ is displayed explicitly by the vertical line in Figs.~(1a) and (1b).}
\label{212Po}
\end{figure}

In Fig.~\ref{212Po}, we study the relation between the cluster-daughter overlap measured by the $\mathfrak{O}$ parameter and the alpha-cluster formation probability $P_\alpha$ by varying the values of $\mu_4$ from $0.95\,Q_\alpha$ to $1.05\,Q_\alpha$. Fig.~(1a) shows the behavior of the effective potential $W(r)$ with the critical radius $r_\text{cluster}=7.43825\text{ fm}$ displayed by the vertical line. As $\mu_4$ becomes larger, the horizontal platform of $W(r)$ within the critical radius $r_\text{cluster}$ becomes higher, and the energy pocket beyond the critical radius $r_\text{cluster}$ becomes deeper. Fig.~(1b) shows the normalized radial wave function $\phi(r)$. As $\mu_4$ becomes larger, more and more parts of the radial wave function $\phi(r)$ are pushed into the outer region $r>r_\text{cluster}$. According to Eq.~\eqref{ACFP}, this indicates that the alpha-cluster formation probability becomes larger and larger. Fig.~(1c) displays the relation between the dimensionless parameter $\mathfrak{O}$ and the alpha-cluster formation probability $P_\alpha$, hinting that there might exist approximately a simple positively-correlated linear relation between $P_\alpha$ and $\mathfrak{O}$. Here, the rms point radii of the alpha particle, the daughter nuclei $^{208}$Pb, and the alpha emitter $^{212}$Po are compiled from the experimental values of their rms charge radii recorded in Ref.~\cite{Angeli:2013epw} by using $R_\text{point}=\sqrt{R^2_\text{ch}-R_o^2}$ with $R_o=0.88$ fm \cite{Brown:2005}. This linear relation allows us to probe the landscape of alpha-cluster formation probabilities by studying that of the cluster-daughter overlap. 

%In summary, in this section we investigate the relation between the cluster-daughter overlap and the alpha-cluster formation probability within the framework of quartetting wave function approach, a simple yet powerful method to study alpha-cluster formation. To measure the degree of the cluster-daughter overlap quantitatively, we introduce a dimensionless $\mathfrak{O}$ parameter as the ratio between the rms intercluster separation and the sum of the rms point radii of the alpha particle and the daughter nucleus. The larger (smaller) the cluster-daughter overlap is, the smaller (larger) the $\mathfrak{O}$ parameter will be. Our calculation shows that, there might exist approximately a simple positively-correlated linear relation between the alpha-cluster formation probability $P_\alpha$ and the dimensionless $\mathfrak{O}$ parameter, which 

\section{THE LANDSCAPE OF ALPHA-CLUSTER FORMATION PROBABILITIES PROBED BY CLUSTER-DAUGHTER OVERLAP}
\label{LANDSCAPE}

We investigate the landscape of alpha-cluster formation probabilities in medium-mass and heavy even-even nuclei based on discussions in Section \ref{ISvsACF}. Explicitly, we will study the landscape of the dimensionless parameter $\mathfrak{O}$ along various isotopic and isotonic chains. Then, according to the linear relation between $P_\alpha$ and $\mathfrak{O}$, the same landscape could also reveal important features of the landscape of alpha-cluster formation probabilities. To extract the $\mathfrak{O}$ parameter from experimental nuclear ground-state charge radii, we make use of the relation \cite{Buck:1975zz,Michel:1988zz,Souza:2015yxa,Souza:2016bpt,Buck:1995zz},
\begin{align}
\braket{R^2}_{\text{ch},P}=\frac{Z_\alpha}{Z_\alpha+Z_d}\braket{R^2}_{\text{ch},\alpha}+\frac{Z_d}{Z_\alpha+Z_d}\braket{R^2}_{\text{ch},d}+\frac{Z_\alpha A^2_d+Z_dA^2_\alpha}{(Z_\alpha+Z_d)(A_\alpha+A_d)^2}\braket{R^2}_i.
\label{InterclusterSeparationMasterFormula}
\end{align}
Here, $\braket{R^2}_{\text{ch},P}^{1/2}$, $\braket{R^2}_{\text{ch},\alpha}^{1/2}$, and $\braket{R^2}_{\text{ch},d}^{1/2}$ are the rms charge radii of the parent nucleus, the alpha particle, the daughter nucleus, while $\braket{R^2}_i^{1/2}$ is the rms separation between the centers of mass of alpha particle and the daughter nucleus. Eq.~\eqref{InterclusterSeparationMasterFormula} assumes that the centers of mass of the alpha particle and the daughter nucleus coincide with their centers of charge, and follows directly from the definition of rms charge radius of the parent nucleus:
\begin{align}
\braket{R^2}_{\text{ch},p}&=\frac{\sum_{i=1}^{Z_d}\braket{(\mathbf{R}_1+\mathbf{r}_i)^2}+\sum_{j=1}^{Z_\alpha}\braket{(\mathbf{R}_2+\mathbf{r}_j)^2}}{Z_\alpha+Z_d}\nonumber\\
&=\frac{Z_d}{Z_\alpha+Z_d}\braket{R^2}_{\text{ch},d}+\frac{Z_\alpha}{Z_\alpha+Z_d}\braket{R^2}_{\text{ch},\alpha}+\frac{Z_d \braket{R_1^2}+Z_\alpha \braket{R_2^2}}{Z_\alpha+Z_d}.
\end{align}
An explanation of the variables here could be found in the caption of Fig.~\ref{Demo}. Taking into account the fact that $\braket{R_1^2}^{1/2}=\frac{A_\alpha}{A_\alpha+A_d}\braket{R^2}^{1/2}_i$ and $\braket{R_2^2}^{1/2}=\frac{A_d}{A_\alpha+A_d}\braket{R^2}^{1/2}_i$, we obtain Eq.~\eqref{InterclusterSeparationMasterFormula}. The experimental values of $\braket{R^2}_{\text{ch},P}^{1/2}$, $\braket{R^2}_{\text{ch},\alpha}^{1/2}$, and $\braket{R^2}_{\text{ch},d}^{1/2}$ could be found in Ref.~\cite{Angeli:2013epw}, from which the value of $R_i\approx\braket{R^2}^{1/2}_i$ is obtained by solving Eq.~\eqref{InterclusterSeparationMasterFormula}. The $\mathfrak{O}$ parameter could then be calculated according to its definition in Eq.~\eqref{DDef}. In this work, we will study the landscape of the $\mathfrak{O}$ parameters of not only stable but also decaying nuclei which are unbound. For the latter case, the definition of the rms radius could be nontrivial. However, the lifetimes of these decaying nuclei are typically very long. For instance, the lifetime of $^{210}$Po is about 140 days, long enough to allow the electron scattering experiment to determine its charge radius practically. As a result, for our current purposes we can talk about the rms radii of these decaying nuclei safely.

\begin{figure}

\centering

\begin{subfigure}[b]{\textwidth}
\centering
\includegraphics[width=0.45\textwidth]{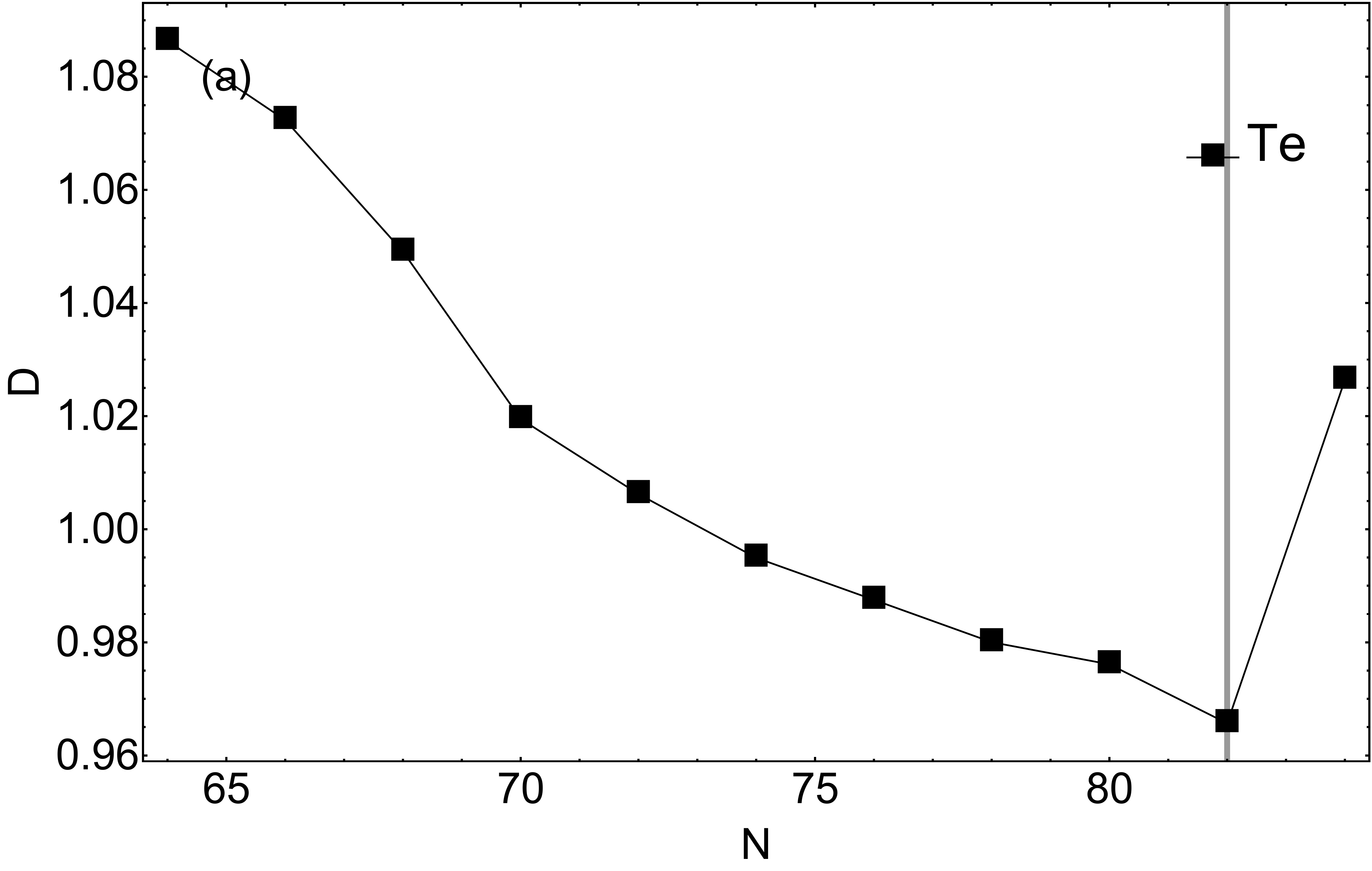}
%\caption*{$\quad\quad\quad$(a)}
\end{subfigure}

\begin{subfigure}[b]{\textwidth}
\centering
\includegraphics[width=0.45\textwidth]{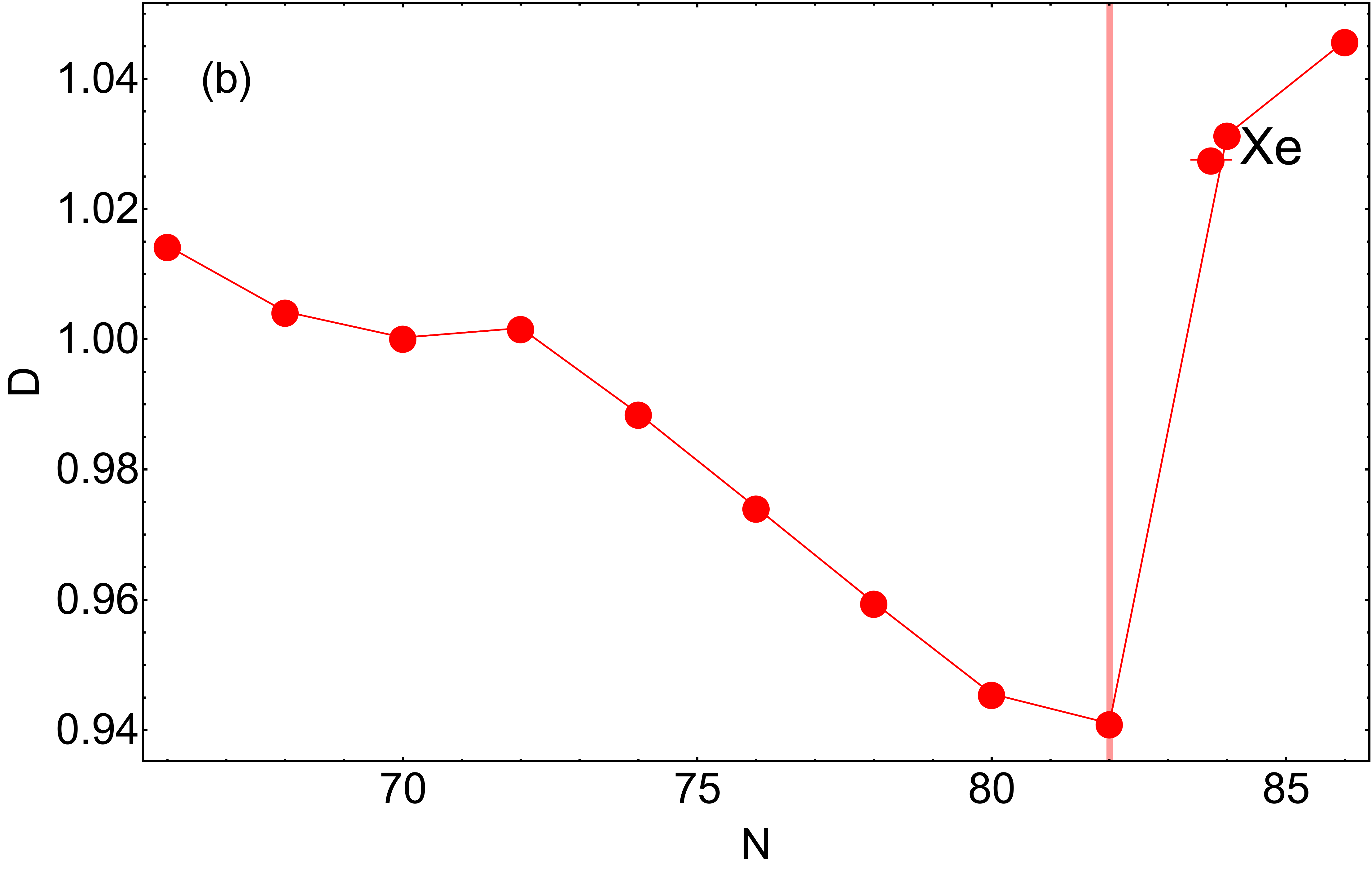}
%\caption*{$\quad\quad\quad$(b)}
\end{subfigure}

\begin{subfigure}[b]{\textwidth}
\centering
\includegraphics[width=0.45\textwidth]{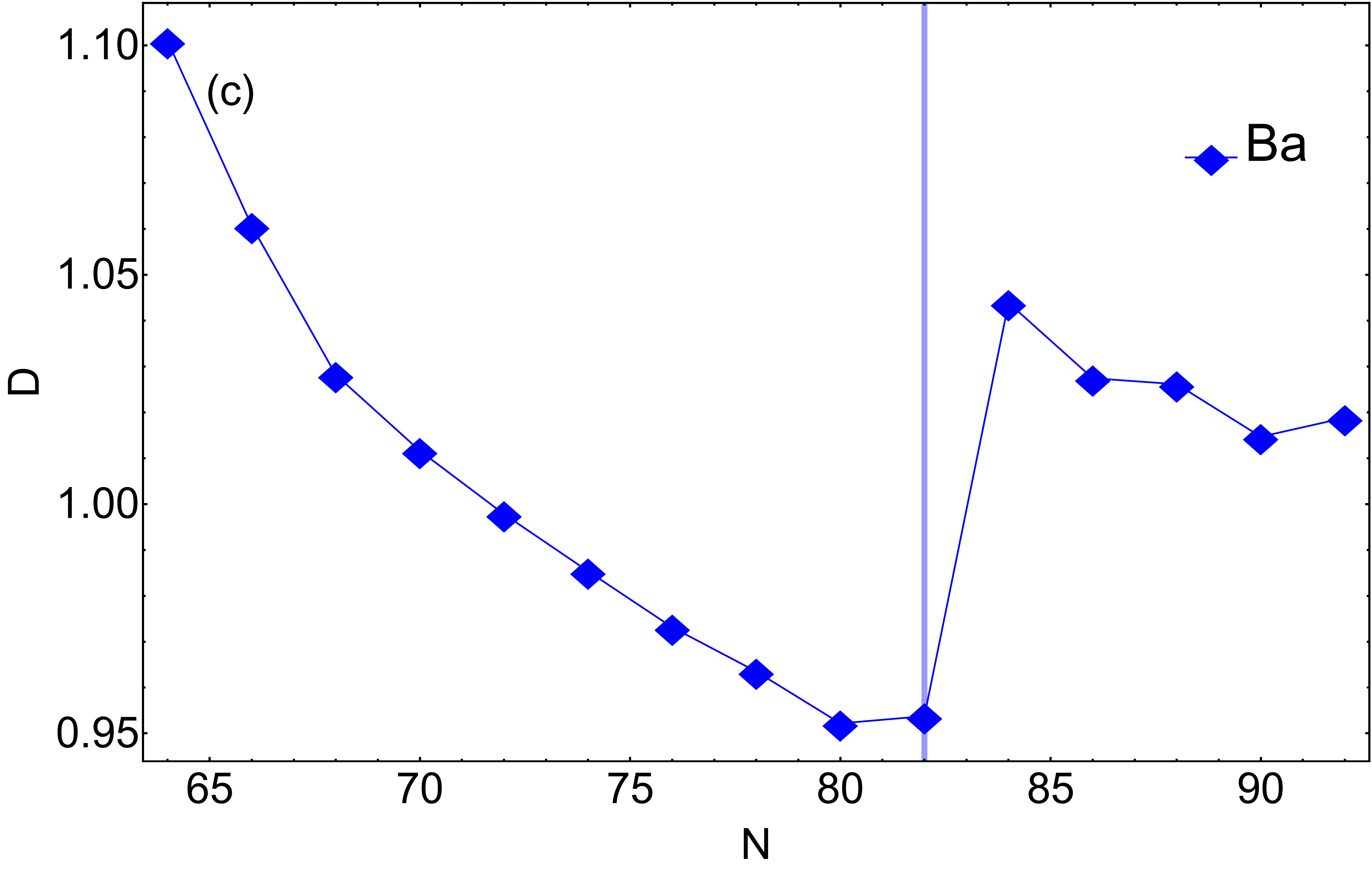}
%\caption*{$\quad\quad\quad$(c)}
\end{subfigure}

\caption{The landscapes of the $\mathfrak{O}$ parameter along the Te, Xe, and Ba isotopes around the $N=82$ shell. The vertical lines denote the $N=82$ shell closure.}
\label{LightN}
\end{figure}

\begin{figure}
\centering
\includegraphics[width=0.45\textwidth]{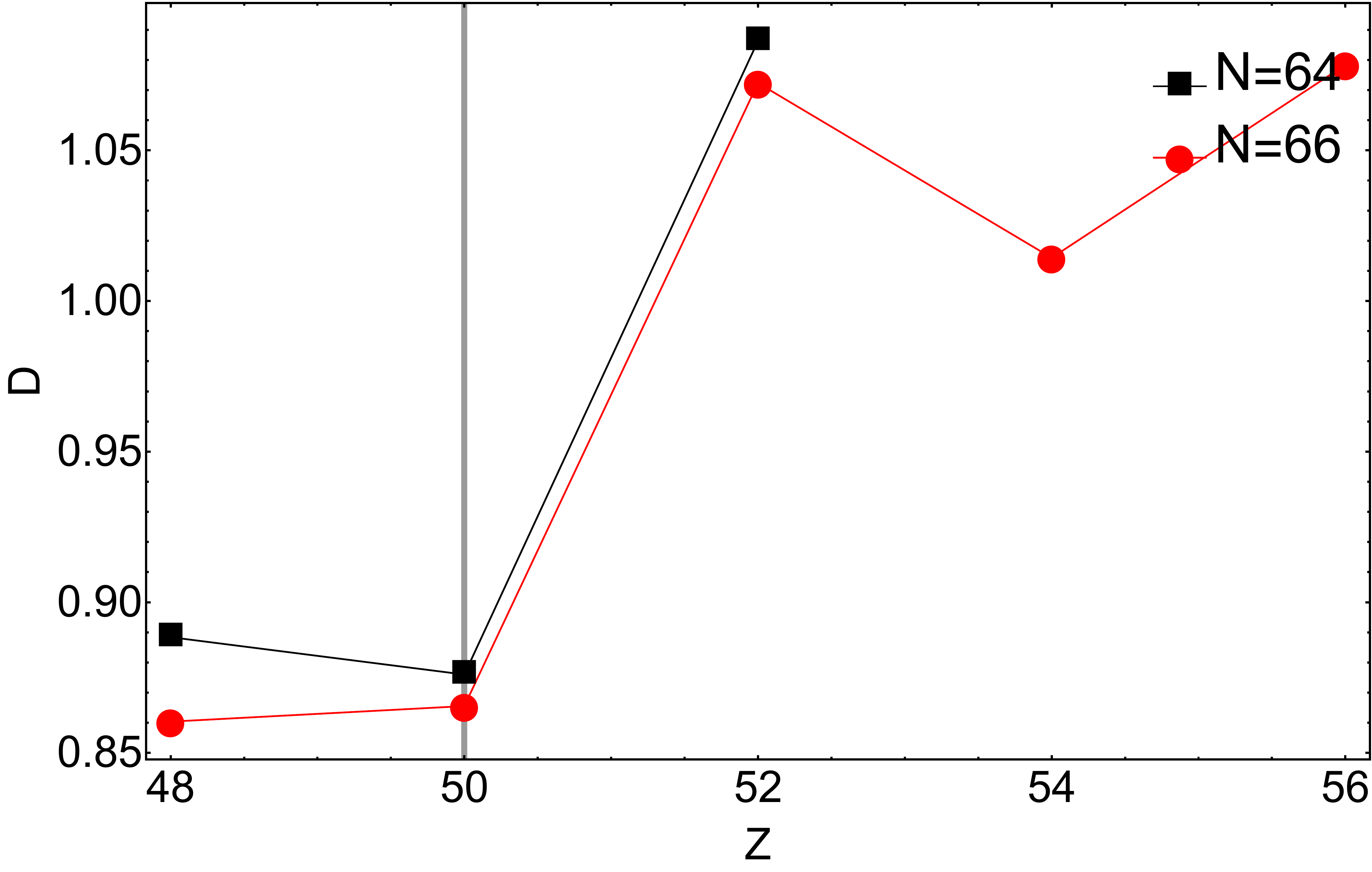}
\caption{The landscape of the $\mathfrak{O}$ parameter along the $N=64$ and $N=66$ isotonic chains around the $Z=50$ shell. The vertical line denotes the $Z=50$ shell closure.}
\label{LightZ}
\end{figure}

We start with the medium-mass nuclei Te, Xe, and Ba, whose $\mathfrak{O}$-parameter landscapes are plotted in Fig.~\ref{LightN}. It is found that, in these three nuclei, the $\mathfrak{O}$ parameter (the alpha-cluster formation probability $P_\alpha$) decreases first and reaches its minimum at around the $N=82$ shell closure. After crossing $N=82$, the $\mathfrak{O}$ parameter (the alpha-cluster formation probability $P_\alpha$) gets suddenly a large increase, manifesting clearly the effect of the $N=82$ shell closure on $P_\alpha$. Such a behavior is consistent with previous studies on heavy nuclei in Ref.~\cite{Batchelder:1997zz,Xu:2017vyt}, which could be viewed as a nontrivial check of the assumption that the alpha-cluster formation probability $P_\alpha$ is approximately proportional to the dimensionless $\mathfrak{O}$ parameter. Compared with Ref.~\cite{Qian:2018bfu}, which studies the alpha-cluster formation probabilities in Te, Xe, and Ba isotopes near the $N=50$ shell based on limited alpha-decay data, the analysis here allows one to get a complementary understanding on how $P_\alpha$ changes around the $N=82$ shell, thanks to the availability of rich experimental data on ground-state charge radii in this region. We also study the effects of the $Z=50$ shell along the $N=64$ and $N=66$ isotonic chains in Fig.~\ref{LightZ}. As expected, a large increase of the $\mathfrak{O}$ parameter (the alpha-cluster formation probability $P_\alpha$) is found from $Z=50$ to $Z=52$. 

\begin{figure}
\centering
\begin{subfigure}[b]{\textwidth}
\centering
\minipage{0.45\textwidth}
  \includegraphics[width=\linewidth]{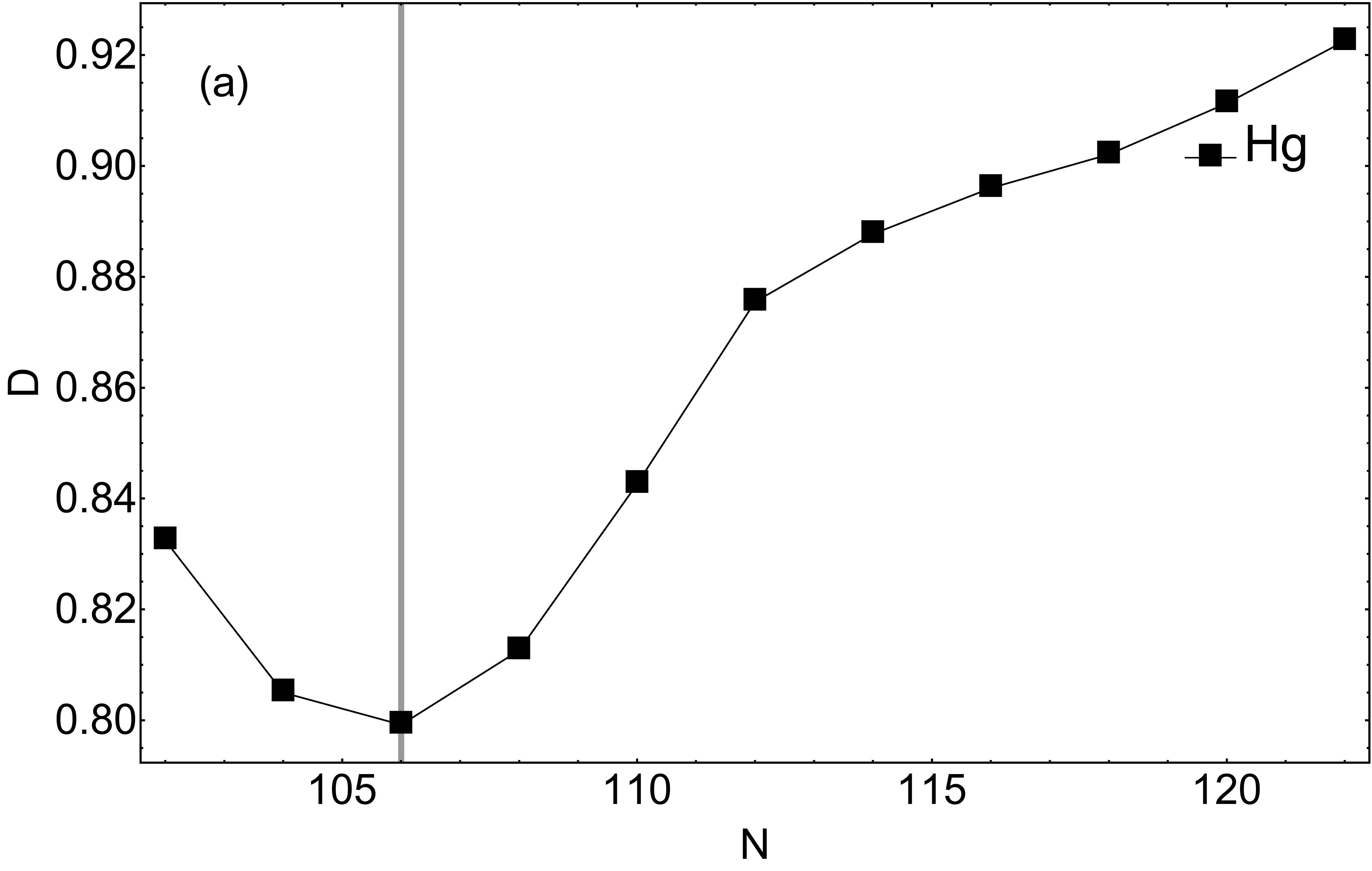}
%  \caption*{$\quad\ \ \ $(a)}
\endminipage\hfill
\minipage{0.45\textwidth}
  \includegraphics[width=\linewidth]{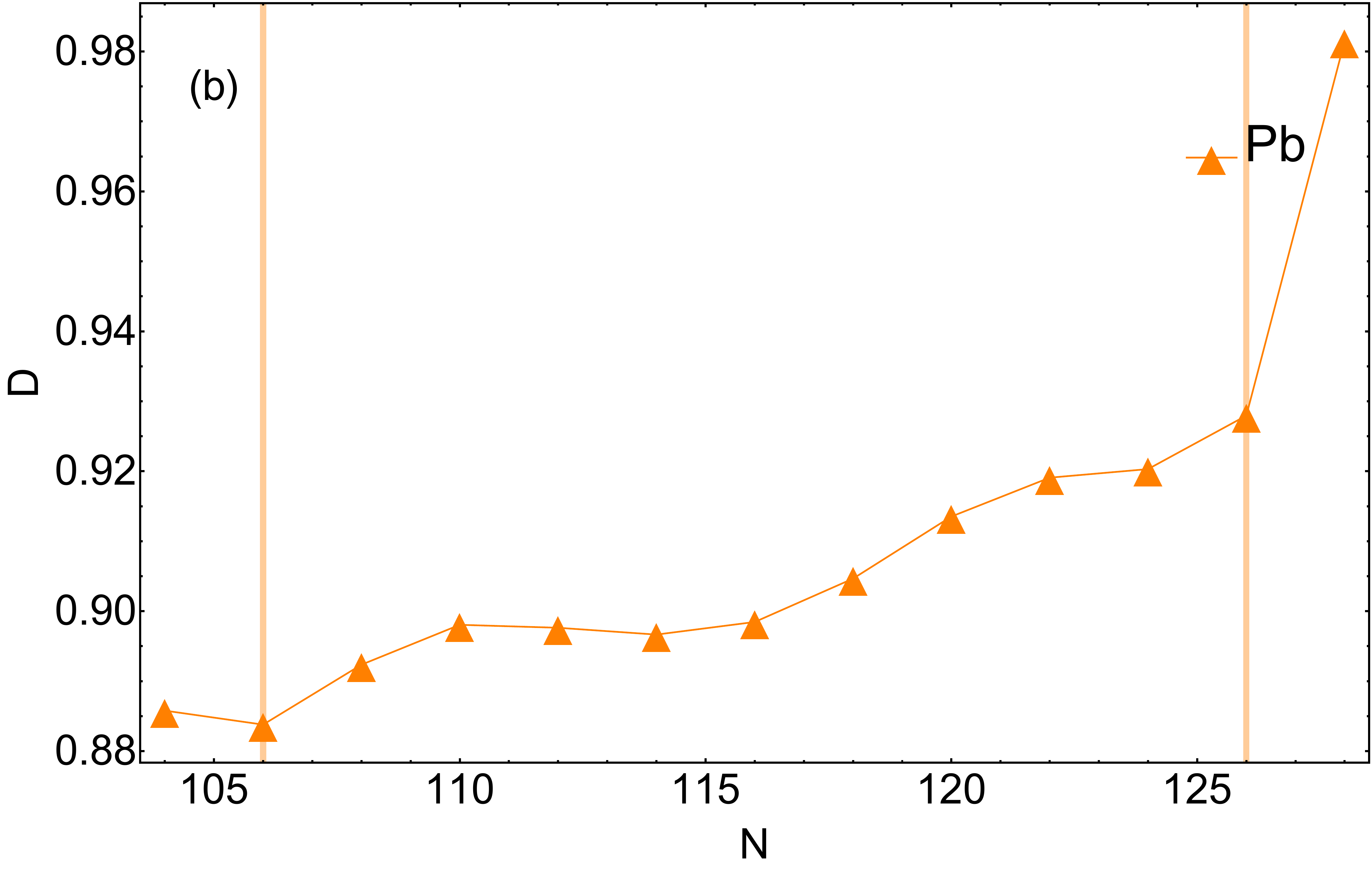}
%  \caption*{$\quad\ \ \ $(b)}
\endminipage
\end{subfigure}

\begin{subfigure}[b]{\textwidth}
\centering
\minipage{0.45\textwidth}
  \includegraphics[width=\linewidth]{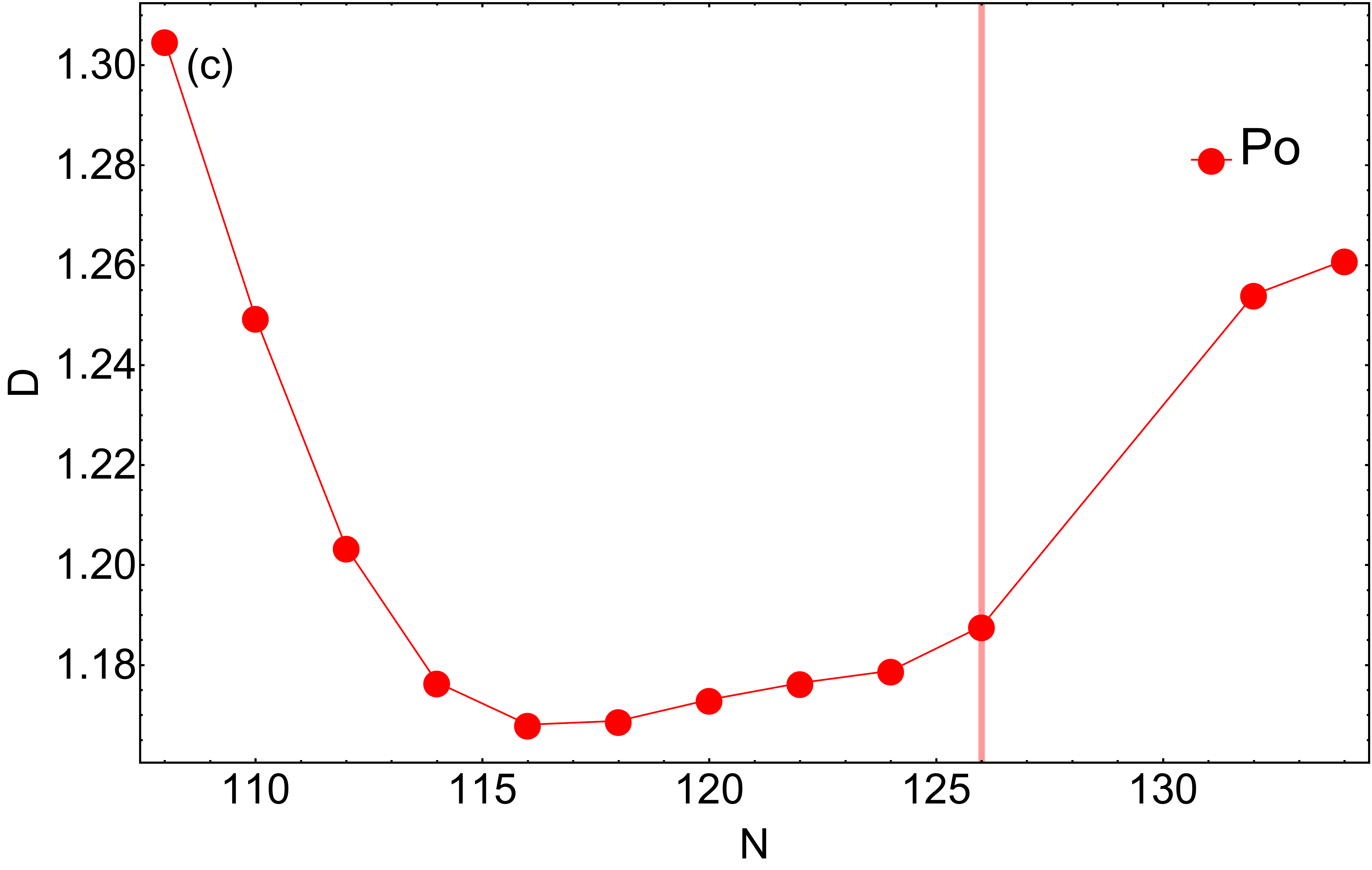}
%  \caption*{$\quad\ \ \ $(c)}
\endminipage\hfill
\minipage{0.45\textwidth}
  \includegraphics[width=\linewidth]{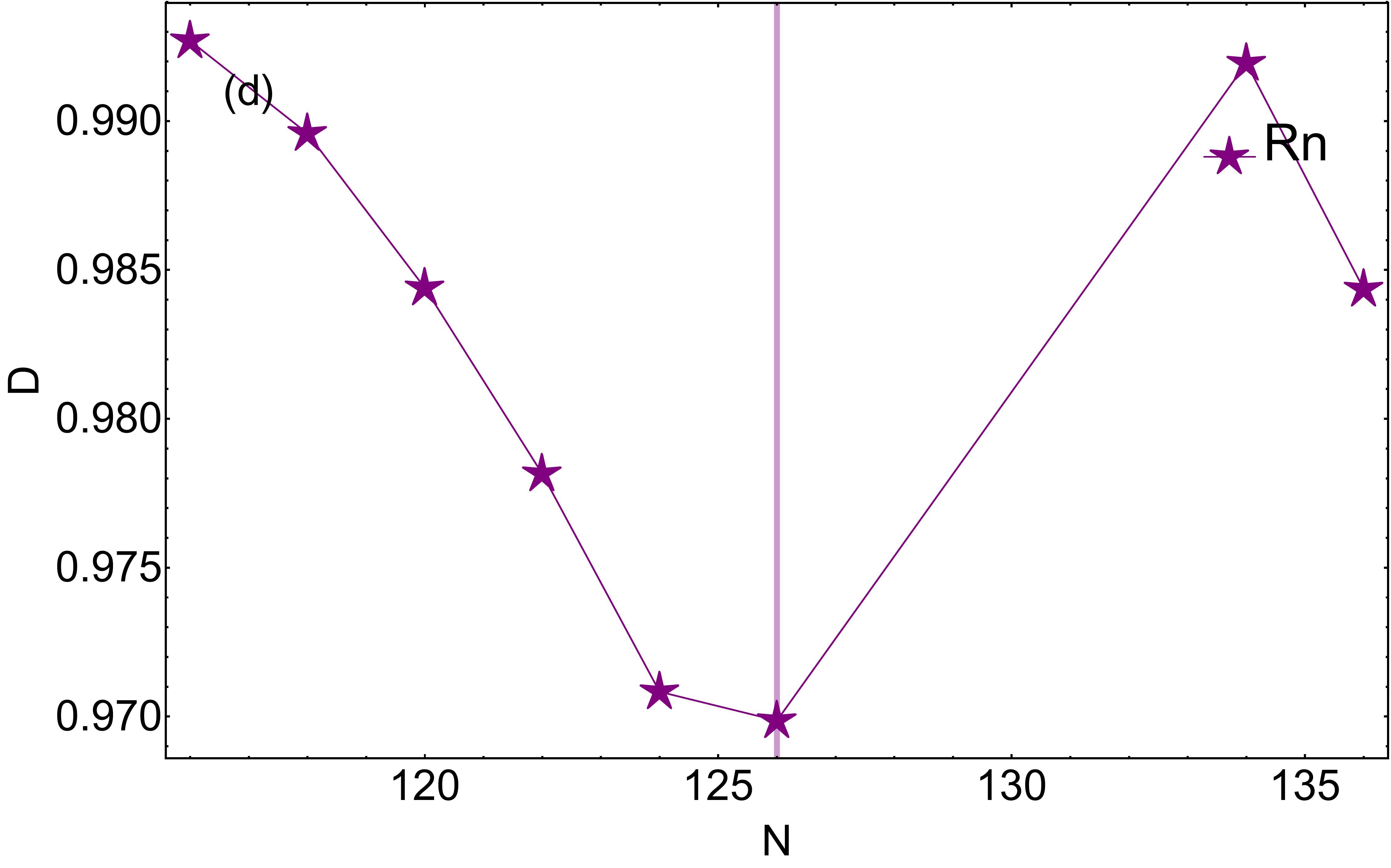}
%  \caption*{$\quad\ \ \ $(d)}
  \endminipage
\end{subfigure}

\begin{subfigure}[b]{\textwidth}
\centering
\minipage{0.45\textwidth}
  \includegraphics[width=\linewidth]{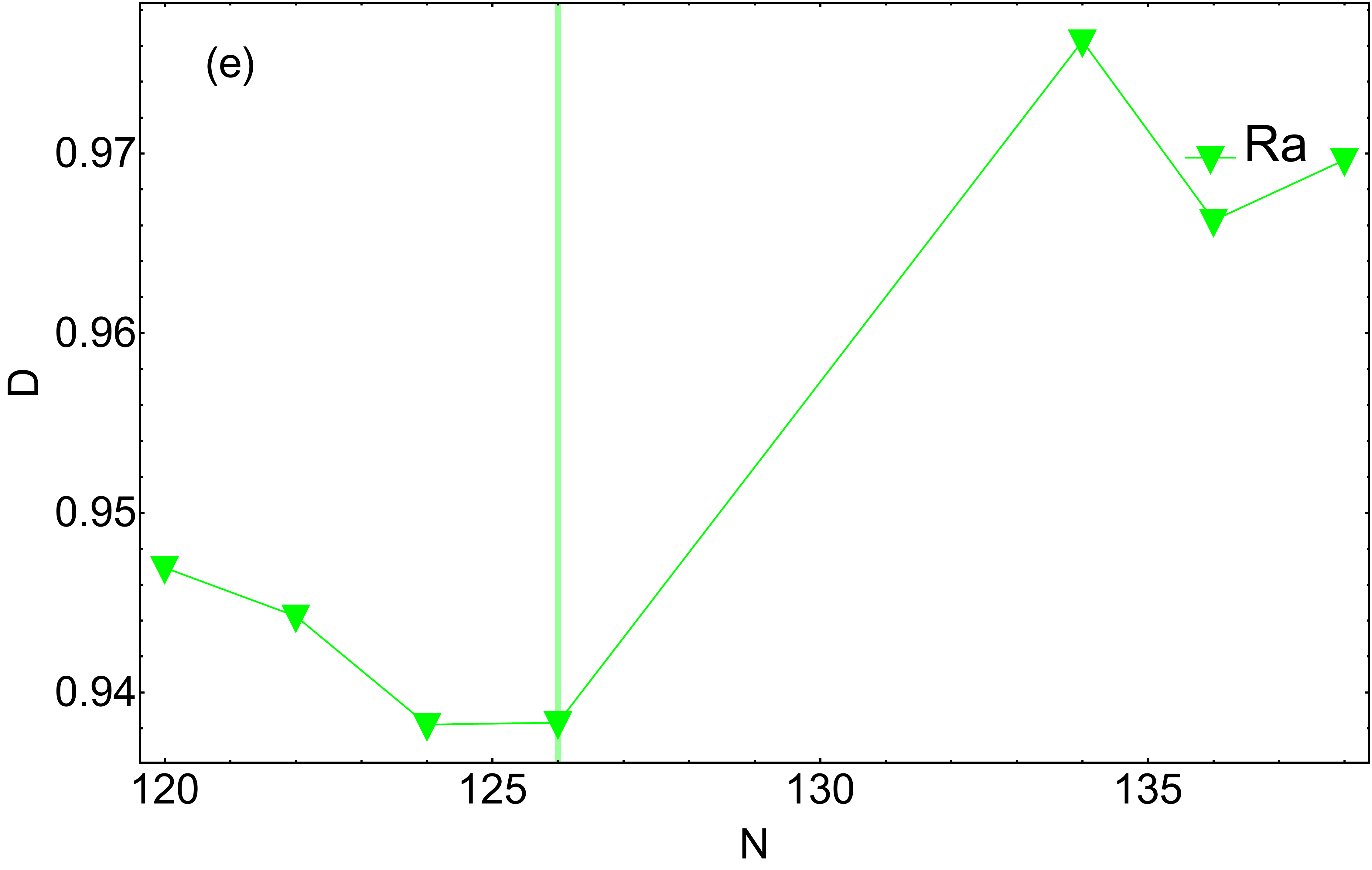}
%  \caption*{$\quad\ \ \ $(c)}
\endminipage
\end{subfigure}

\caption{The landscapes of the $\mathfrak{O}$ parameter along the Hg, Pb, Po, Rn, and Ra isotopes around the $N=126$ shell. The vertical line in Fig.~(4a) denotes the $N=106$ subshell. The two vertical lines in Fig.~(4b) denote the $N=106$ subshell and the $N=126$ shell, respectively. The vertical lines in Fig.~(4c)-(4e) denote the $N=126$ major shell.}
\label{HeavyN}
\end{figure}

\begin{figure}
\centering
\includegraphics[width=0.45\textwidth]{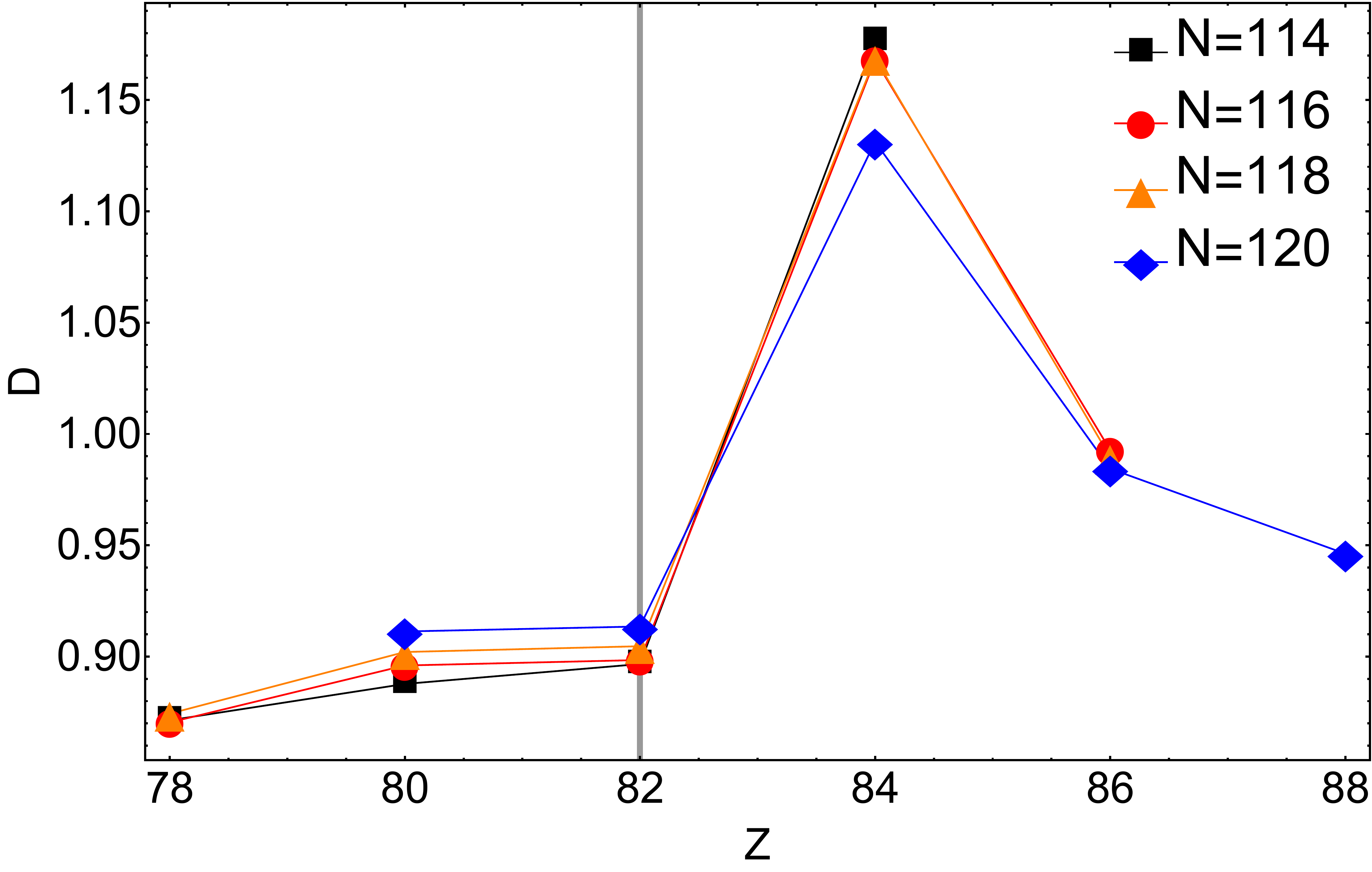}
\caption{The landscape of the $\mathfrak{O}$ parameters along the $N=114$, $116$, $118$, and $120$ isotonic chains near the $Z=82$ shell. The vertical line denotes the $Z=82$ shell closure.}
\label{HeavyZ}
\end{figure}

As mentioned before, lots of works have been devoted to study the alpha-cluster formation probabilities of heavy nuclei by exploiting alpha-decay data. Here, we would like to provide a complementary calculation by using experimental data of nuclear rms radii. The $\mathfrak{O}$-parameter landscapes of the heavy nuclei Hg, Pb, Po, Rn, and Ra around the $N=126$ major shell could be found in Fig.~\ref{HeavyN}. For the Po, Rn, and Ra isotopic chains in Fig.~(5c), (5d) and (5e), the $\mathfrak{O}$ parameters decrease first and get a large increase across the $N=126$ shell, consistent with previous studies based on alpha-decay data \cite{Batchelder:1997zz,Xu:2017vyt}. In Fig.~(5a), we plot the $\mathfrak{O}$ parameters for the Hg isotopes in the region $N=102-122$. It is interesting to see that the $N=106$ subshell effect is made manifest in this plot, suggesting that the alpha-cluster formation probability gets a large increase when crossing the $N=106$ subshell along the Hg isotopic chain. For the Pb isotopic chain shown in Fig.~(5b), it displays both the $N=106$ subshell and the $N=126$ shell structures, with the $\mathfrak{O}$ parameters (the alpha-cluster formation probabilities $P_\alpha$) getting a large increase at both $N=106$ and $N=126$. We study further the $Z=82$ shell effects along the $N=114$, $116$, $118$, and $120$ isotonic chains in Fig.~\ref{HeavyZ}. Similar to the medium-mass nuclei, a large increase of the $\mathfrak{O}$ parameter (the alpha-cluster formation probability $P_\alpha$) is found from $Z=82$ to $Z=84$, consistent with Refs.~\cite{Andreyev:2013iwa,Xu:2017vyt}.

\section{CONCLUSIONS}
\label{CONCLUSIONS}

In this note, we study the possibilities to use the cluster-daughter overlap as a new probe of alpha-cluster formation in medium-mass and heavy even-even nuclei. Quantitatively, the cluster-daughter overlap is measured by a dimensionless $\mathfrak{O}$ parameter which is the ratio between the rms intercluster separation and the sum of the rms point radii of the alpha particle and the daughter nucleus. The larger (smaller) the cluster-daughter overlap is, the smaller (larger) the $\mathfrak{O}$ parameter will be. Within the quartetting wave function approach, we show that there might exist approximately a positively-correlated linear relation between the alpha-cluster formation probability $P_\alpha$ and the $\mathfrak{O}$ parameter, which enables one to probe the landscape of alpha-cluster formation probabilities by studying instead the landscape of the $\mathfrak{O}$ parameter. By exploiting the experimental values of the ground-state charge radii of various medium-mass and heavy nuclei, we plot the landscape of the $\mathfrak{O}$ parameters. The general trends of the alpha-cluster formation probabilities derived thereby are consistent with previous studies and natural theoretical expectations. The shell effects at $N=82$, $126$ and $Z=50$, $82$ are identified explicitly in terms of the cluster-daughter overlap, as well as interesting subshell effects at $N=106$ along the Hg and Pb isotopic chains. Compared with previous studies based on alpha-decay data, the new probe could be used in cases where the target nucleus is stable against the alpha decay or the alpha-decay data are temporarily unavailable. The study here could be helpful for future theoretical and experimental studies on alpha-cluster formation in medium-mass and heavy regions, as well as providing extra motivations to carry out new experiments to measure charge radii in the medium-mass and heavy regions precisely. In this work, we assume both the alpha cluster and the daughter nucleus to be spherical. It is also important to consider further the impact of nuclear deformations, which according to previous studies such as Ref.~\cite{Ren:1987,Ren:1988} would enhance the alpha-cluster formation probability. Hopefully, this would be the topic of our future studies.

\begin{acknowledgments} 
D.~B.~would like to thank the organizers of the Fourth International Workshop on ``State of the Art in Nuclear Cluster Physics'', May 13-18, at Galveston, Texas, USA, for the oral talk invitation, during which this work is prepared. This work is supported by the National Natural Science Foundation of China (Grant No.~11535004, 11761161001, 11375086, 11120101005, 11175085 and 11235001), by the National Key R\&D Program of China (Contract No.~2018YFA0404403, 2016YFE0129300), and by the Science and Technology Development Fund of Macau under Grant No.~008/2017/AFJ.
\end{acknowledgments}


\begin{thebibliography}{999}

\bibitem{Beck:2010}
C.~Beck ed., \emph{Clusters in Nuclei}, Lecture Notes in Physics, Vol. 1 (Springer-Verlag, Berlin and Heidelberg, 2010).

\bibitem{Beck:2012}
C.~Beck ed., \emph{Clusters in Nuclei}, Lecture Notes in Physics, Vol. 2 (Springer-Verlag, Berlin and Heidelberg, 2012).

\bibitem{Beck:2014}
C.~Beck ed., \emph{Clusters in Nuclei}, Lecture Notes in Physics, Vol. 3 (Springer-Verlag, Berlin and Heidelberg, 2014).

\bibitem{Horiuchi:2012}
H.~Horiuchi, K.~Ikeda, and K.~Kato, 
``Recent Developments in Nuclear Cluster Physics,''
Prog.\ Theor.\ Phys.\ Suppl.\ {\bf 192}, 1 (2012).

%\cite{Freer:2017gip}
\bibitem{Freer:2017gip} 
  M.~Freer, H.~Horiuchi, Y.~Kanada-En'yo, D.~Lee and U.~G.~Meissner,
  ``Microscopic Clustering in Light Nuclei,''
  arXiv:1705.06192 [nucl-th].
  %%CITATION = ARXIV:1705.06192;%%
  %16 citations counted in INSPIRE as of 29 May 2018
  
       \bibitem{Delion:2010}
 D.~S.~Delion, \emph{Theory of Particle and Cluster Emission} (Springer-Verlag, Berlin, 2010).

%\cite{Wheeler:1937zza}
\bibitem{Wheeler:1937zza} 
  J.~A.~Wheeler,
  ``Molecular Viewpoints in Nuclear Structure,''
  Phys.\ Rev.\  {\bf 52}, 1083 (1937).
  %doi:10.1103/PhysRev.52.1083
  %%CITATION = doi:10.1103/PhysRev.52.1083;%%
  %229 citations counted in INSPIRE as of 29 May 2018
  
  \bibitem{Hafstad:1938}
  L.~R.~Hafstad and E.~Teller, 
  ¡°The alpha-particle model of the nucleus,¡± 
  Phys.\ Rev.\ {\bf 54}, 681 (1938).
  
  %\cite{Wheeler:1937zz}
\bibitem{Wheeler:1937zz} 
  J.~A.~Wheeler,
  ``On the Mathematical Description of Light Nuclei by the Method of Resonating Group Structure,''
  Phys.\ Rev.\  {\bf 52}, 1107 (1937).
  %doi:10.1103/PhysRev.52.1107
  %%CITATION = doi:10.1103/PhysRev.52.1107;%%
  %195 citations counted in INSPIRE as of 29 May 2018
  
  \bibitem{Brink:1966}
  D.~M.~Brink, 
  ``Alpha cluster model,''
  in Proceedings of the International School of Physics Enrico Fermi, Varenna Course 36, p.~247 (1966).
  
%\cite{Saito:1969zz}
\bibitem{Saito:1969zz} 
  S.~Saito,
  ``Interaction between clusters and Pauli principle,''
  Prog.\ Theor.\ Phys.\  {\bf 41}, 705 (1969).
  %doi:10.1143/PTP.41.705
  %%CITATION = doi:10.1143/PTP.41.705;%%
  %254 citations counted in INSPIRE as of 29 May 2018
  
  %\cite{Kanada-Enyo:2001yji}
\bibitem{Kanada-Enyo:2001yji} 
  Y.~Kanada-En'yo and H.~Horiuchi,
  ``Structure of light unstable nuclei studied with antisymmetrized molecular dynamics,''
  Prog.\ Theor.\ Phys.\ Suppl.\  {\bf 142}, 205 (2001)
  %doi:10.1143/PTPS.142.205
  [nucl-th/0107044].
  %%CITATION = doi:10.1143/PTPS.142.205;%%
  %123 citations counted in INSPIRE as of 29 May 2018
  
  %\cite{Tohsaki:2001an}
\bibitem{Tohsaki:2001an} 
  A.~Tohsaki, H.~Horiuchi, P.~Schuck and G.~R\"opke,
  ``Alpha cluster condensation in $^{12}$C and $^{16}$O,''
  Phys.\ Rev.\ Lett.\  {\bf 87}, 192501 (2001)
  %doi:10.1103/PhysRevLett.87.192501
  [nucl-th/0110014].
  %%CITATION = doi:10.1103/PhysRevLett.87.192501;%%
  %357 citations counted in INSPIRE as of 29 May 2018
  
  %\cite{Funaki:2008gb}
\bibitem{Funaki:2008gb} 
  Y.~Funaki, T.~Yamada, H.~Horiuchi, G.~R\"opke, P.~Schuck and A.~Tohsaki,
  ``Alpha-particle condensation in $^{16}$O via a full four-body OCM calculation,''
  Phys.\ Rev.\ Lett.\  {\bf 101}, 082502 (2008)
  %doi:10.1103/PhysRevLett.101.082502
  [arXiv:0802.3246 [nucl-th]].
  %%CITATION = doi:10.1103/PhysRevLett.101.082502;%%
  %95 citations counted in INSPIRE as of 29 May 2018
  
  %\cite{Funaki:2009fc}
\bibitem{Funaki:2009fc} 
  Y.~Funaki, H.~Horiuchi, W.~von Oertzen, G.~R\"opke, P.~Schuck, A.~Tohsaki and T.~Yamada,
  ``Concepts of nuclear alpha-particle condensation,''
  Phys.\ Rev.\ C {\bf 80}, 064326 (2009)
  %doi:10.1103/PhysRevC.80.064326
  [arXiv:0912.2934 [nucl-th]].
  %%CITATION = doi:10.1103/PhysRevC.80.064326;%%
  %72 citations counted in INSPIRE as of 29 May 2018
  
  %\cite{Yamada:2003cz}
\bibitem{Yamada:2003cz} 
  T.~Yamada and P.~Schuck,
  ``Dilute multi-$\alpha$ cluster states in nuclei,''
  Phys.\ Rev.\ C {\bf 69}, 024309 (2004)
  %doi:10.1103/PhysRevC.69.024309
  [nucl-th/0310077].
  %%CITATION = doi:10.1103/PhysRevC.69.024309;%%
  %83 citations counted in INSPIRE as of 29 May 2018
  
  %\cite{Bai:2018gqt}
\bibitem{Bai:2018gqt} 
  D.~Bai and Z.~Ren,
  ``Repulsive Four-Body Interactions of $\alpha$ Particles and Quasi-Stable Nuclear $\alpha$-Particle Condensates in Heavy Self-Conjugate Nuclei,''
  Phys.\ Rev.\ C {\bf 97}, 054301 (2018)
  %doi:10.1103/PhysRevC.97.054301
  [arXiv:1804.05992 [nucl-th]].
  %%CITATION = doi:10.1103/PhysRevC.97.054301;%%
  
  \bibitem{Rutherford:1899}
  E.~Rutherford, 
  ``Uranium radiation and the electrical conduction produced by it,''
  Philos.\ Mag.\ {\bf 47}, 109 (1899).
  
  %\cite{Delion:2018rrl}
\bibitem{Delion:2018rrl} 
  D.~S.~Delion, Z.~Ren, A.~Dumitrescu and D.~Ni,
  ``Coupled channels description of the $\alpha$-decay fine structure,''
  J.\ Phys.\ G {\bf 45}, 053001 (2018).
  %doi:10.1088/1361-6471/aaac52
  %%CITATION = doi:10.1088/1361-6471/aaac52;%%
  %1 citations counted in INSPIRE as of 29 May 2018
  
  %\cite{Buck:1992zz}
\bibitem{Buck:1992zz} 
  B.~Buck, A.~C.~Merchant and S.~M.~Perez,
  ``$\alpha$ decay calculations with a realistic potential,''
  Phys.\ Rev.\ C {\bf 45}, 2247 (1992).
  %doi:10.1103/PhysRevC.45.2247
  %%CITATION = doi:10.1103/PhysRevC.45.2247;%%
  %107 citations counted in INSPIRE as of 29 May 2018

  %\cite{Xu:2005ukj}
\bibitem{Xu:2005ukj} 
  C.~Xu and Z.~Ren,
  ``Systematical calculation of $\alpha$ decay half-lives by density-dependent cluster model,''
  Nucl.\ Phys.\ A {\bf 753}, 174 (2005).
  %doi:10.1016/j.nuclphysa.2005.02.125
  %%CITATION = doi:10.1016/j.nuclphysa.2005.02.125;%%
  %75 citations counted in INSPIRE as of 29 May 2018
  
  %\cite{Delion:2017ozx}
\bibitem{Delion:2017ozx} 
  D.~S.~Delion and S.~A.~Ghinescu,
  ``Geiger-Nuttall Law for Nuclei in Strong Electromagnetic Fields,''
  Phys.\ Rev.\ Lett.\  {\bf 119}, 202501 (2017).
  %doi:10.1103/PhysRevLett.119.202501
  %%CITATION = doi:10.1103/PhysRevLett.119.202501;%%
  %2 citations counted in INSPIRE as of 29 May 2018

%\cite{Bai:2018adq}
\bibitem{Bai:2018adq} 
  D.~Bai, D.~Deng and Z.~Ren,
  ``Charged Particle Emissions in High-Frequency Alternative Electric Fields,''
  Nucl.\ Phys.\ A {\bf 976}, 23 (2018)
  %doi:10.1016/j.nuclphysa.2018.05.004
  [arXiv:1805.02379 [nucl-th]].
  %%CITATION = doi:10.1016/j.nuclphysa.2018.05.004;%%
  
  %\cite{Zhongzhou:1987zz}
\bibitem{Zhongzhou:1987zz} 
  Z.~Ren and G.~Xu,
  ``Reduced alpha transfer rates in a schematic model,''
  Phys.\ Rev.\ C {\bf 36}, 456 (1987).
 % doi:10.1103/PhysRevC.36.456
  %%CITATION = doi:10.1103/PhysRevC.36.456;%%
  %43 citations counted in INSPIRE as of 29 May 2018
  
  %\cite{Zhongzhou:1988zz}
\bibitem{Zhongzhou:1988zz} 
  Z.~Ren and G.~Xu,
  ``Evidence of alpha correlation from binding energies in medium and heavy nuclei,''
  Phys.\ Rev.\ C {\bf 38}, 1078 (1988).
  %doi:10.1103/PhysRevC.38.1078
  %%CITATION = doi:10.1103/PhysRevC.38.1078;%%
  %27 citations counted in INSPIRE as of 29 May 2018
  
  %\cite{Hatsukawa:1990zz}
\bibitem{Hatsukawa:1990zz} 
  Y.~Hatsukawa, H.~Nakahara and D.~C.~Hoffman,
  ``Systematics of alpha decay half-lives,''
  Phys.\ Rev.\ C {\bf 42}, 674 (1990).
  %doi:10.1103/PhysRevC.42.674
  %%CITATION = doi:10.1103/PhysRevC.42.674;%%
  %33 citations counted in INSPIRE as of 29 May 2018

  %\cite{Brown:1992rg}
\bibitem{Brown:1992rg} 
  B.~A.~Brown,
  ``Simple relation for alpha decay half-lives,''
  Phys.\ Rev.\ C {\bf 46}, 811 (1992).
  %doi:10.1103/PhysRevC.46.811
  %%CITATION = doi:10.1103/PhysRevC.46.811;%%
  %93 citations counted in INSPIRE as of 29 May 2018
  
  %\cite{Batchelder:1997zz}
\bibitem{Batchelder:1997zz} 
  J.~C.~Batchelder {\it et al.},
  ``$\alpha$-decay properties of $^{190}$Po and the identification of $^{191}$Po,''
  Phys.\ Rev.\ C {\bf 55}, 2142 (1997).
  %doi:10.1103/PhysRevC.55.R2142
  %%CITATION = doi:10.1103/PhysRevC.55.R2142;%%
  %10 citations counted in INSPIRE as of 29 May 2018
  
  %\cite{Delion:2004wv}
\bibitem{Delion:2004wv} 
  D.~S.~Delion, A.~Sandulescu and W.~Greiner,
  ``Evidence for alpha clustering in heavy and superheavy nuclei,''
  Phys.\ Rev.\ C {\bf 69}, 044318 (2004).
  %doi:10.1103/PhysRevC.69.044318
  %%CITATION = doi:10.1103/PhysRevC.69.044318;%%
  %42 citations counted in INSPIRE as of 29 May 2018
  
  %\cite{Xu:2006fq}
\bibitem{Xu:2006fq} 
  C.~Xu and Z.~Ren,
  ``Global calculation of alpha-decay half-lives with a deformed density-dependent cluster model,''
  Phys.\ Rev.\ C {\bf 74}, 014304 (2006).
  %doi:10.1103/PhysRevC.74.014304
  %%CITATION = doi:10.1103/PhysRevC.74.014304;%%
  %96 citations counted in INSPIRE as of 29 May 2018
  
  %\cite{Denisov:2009ng}
\bibitem{Denisov:2009ng} 
  V.~Y.~Denisov and A.~A.~Khudenko,
  ``$\alpha$-decay half-lives: Empirical relations,''
  Phys.\ Rev.\ C {\bf 79}, 054614 (2009)
  Erratum: [Phys.\ Rev.\ C {\bf 82}, 059901 (2010)]
  %doi:10.1103/PhysRevC.82.059901, 10.1103/PhysRevC.79.054614
  [arXiv:0902.0677 [nucl-th]].
  %%CITATION = doi:10.1103/PhysRevC.82.059901, 10.1103/PhysRevC.79.054614;%%
  %63 citations counted in INSPIRE as of 29 May 2018
  
  %\cite{Zhang:2009zzs}
\bibitem{Zhang:2009zzs} 
  G.~L.~Zhang, X.~Y.~Le and H.~Q.~Zhang,
  ``Determination of alpha preformation for heavy nuclei,''
  Nucl.\ Phys.\ A {\bf 823}, 16 (2009).
  %doi:10.1016/j.nuclphysa.2009.03.005
  %%CITATION = doi:10.1016/j.nuclphysa.2009.03.005;%%
  %19 citations counted in INSPIRE as of 29 May 2018
  
  %\cite{Zhang:2008ct}
\bibitem{Zhang:2008ct} 
  H.~F.~Zhang and G.~Royer,
  ``$\alpha$ particle preformation in heavy nuclei and penetration probability,''
  Phys.\ Rev.\ C {\bf 77}, 054318 (2008)
  %doi:10.1103/PhysRevC.77.054318
  [arXiv:0805.4523 [nucl-ex]].
  %%CITATION = doi:10.1103/PhysRevC.77.054318;%%
  %64 citations counted in INSPIRE as of 29 May 2018

%\cite{Andreyev:2013iwa}
\bibitem{Andreyev:2013iwa} 
  A.~N.~Andreyev {\it et al.},
  ``Signatures of the $Z=82$ Shell Closure in $\alpha$-Decay Process,''
  Phys.\ Rev.\ Lett.\  {\bf 110}, 242502 (2013).
  %doi:10.1103/PhysRevLett.110.242502
  %%CITATION = doi:10.1103/PhysRevLett.110.242502;%%
  %35 citations counted in INSPIRE as of 29 May 2018
  
  %\cite{Qi:2014ska}
\bibitem{Qi:2014ska} 
  C.~Qi, A.~N.~Andreyev, M.~Huyse, R.~J.~Liotta, P.~Van Duppen and R.~Wyss,
  ``On the validity of the Geiger-Nuttall alpha-decay law and its microscopic basis,''
  Phys.\ Lett.\ B {\bf 734}, 203 (2014)
  %doi:10.1016/j.physletb.2014.05.066
  [arXiv:1405.5633 [nucl-th]].
  %%CITATION = doi:10.1016/j.physletb.2014.05.066;%%
  %25 citations counted in INSPIRE as of 29 May 2018
  
  %\cite{Khuyagbaatar:2015hjj}
\bibitem{Khuyagbaatar:2015hjj} 
  J.~Khuyagbaatar {\it et al.},
  ``New Short-Lived Isotope $^{221}$U and the Mass Surface Near N=126,''
  Phys.\ Rev.\ Lett.\  {\bf 115}, 242502 (2015).
  %doi:10.1103/PhysRevLett.115.242502
  %%CITATION = doi:10.1103/PhysRevLett.115.242502;%%
  %16 citations counted in INSPIRE as of 29 May 2018
  
  %\cite{Qian:2018dyz}
\bibitem{Qian:2018dyz} 
  Y.~Qian and Z.~Ren,
  ``New insight into $\alpha$ clustering of heavy nuclei via their $\alpha$ decay,''
  Phys.\ Lett.\ B {\bf 777}, 298 (2018).
  %doi:10.1016/j.physletb.2017.12.046
  %%CITATION = doi:10.1016/j.physletb.2017.12.046;%%
  %1 citations counted in INSPIRE as of 29 May 2018
  
  %\cite{Souza:2016bpt}
\bibitem{Souza:2016bpt} 
  M.~A.~Souza and H.~Miyake,
  ``Search for $\alpha$ + core states in even-even Cr isotopes,''
  Eur.\ Phys.\ J.\ A {\bf 53}, 146 (2017)
  %doi:10.1140/epja/i2017-12339-9
  [arXiv:1612.06805 [nucl-th]].
  %%CITATION = doi:10.1140/epja/i2017-12339-9;%%
  %1 citations counted in INSPIRE as of 29 May 2018
  
  %\cite{Mohr:2017zot}
\bibitem{Mohr:2017zot} 
  P.~Mohr,
  ``$\alpha$-cluster states in $^{46,54}$Cr from double-folding potentials,''
  Eur.\ Phys.\ J.\ A {\bf 53}, 209 (2017)
  %doi:10.1140/epja/i2017-12410-7
  [arXiv:1710.02790 [nucl-th]].
  %%CITATION = doi:10.1140/epja/i2017-12410-7;%%
  
  %\cite{Souza:2015yxa}
\bibitem{Souza:2015yxa} 
  M.~A.~Souza and H.~Miyake,
  ``$\alpha$-cluster structure in even-even nuclei around $^{94}$Mo,''
  Phys.\ Rev.\ C {\bf 91}, 034320 (2015)
  %doi:10.1103/PhysRevC.91.034320
  [arXiv:1502.06042 [nucl-th]].
  %%CITATION = doi:10.1103/PhysRevC.91.034320;%%
  %5 citations counted in INSPIRE as of 29 May 2018
  
  %\cite{Ohkubo:1995zz}
\bibitem{Ohkubo:1995zz} 
  S.~Ohkubo,
  ``Alpha Clustering and Structure of $^{94}$Mo and $^{212}$Po,''
  Phys.\ Rev.\ Lett.\  {\bf 74}, 2176 (1995).
  %doi:10.1103/PhysRevLett.74.2176
  %%CITATION = doi:10.1103/PhysRevLett.74.2176;%%
  %36 citations counted in INSPIRE as of 29 May 2018
  
  %\cite{Buck:1995zza}
\bibitem{Buck:1995zza} 
  B.~Buck, A.~C.~Merchant and S.~M.~Perez,
  ``Systematics of alpha-cluster states above double shell closures,''
  Phys.\ Rev.\ C {\bf 51}, 559 (1995).
  %doi:10.1103/PhysRevC.51.559
  %%CITATION = doi:10.1103/PhysRevC.51.559;%%
  %57 citations counted in INSPIRE as of 29 May 2018
  
  %\cite{Michel:2000nw}
\bibitem{Michel:2000nw} 
  F.~Michel, G.~Reidemeister and S.~Ohkubo,
  ``Unexpected transparency in low-energy $^{90}$Zr($\alpha$, $\alpha_0$) scattering and alpha cluster structure in $^{94}$Mo,''
  Phys.\ Rev.\ C {\bf 61}, 041601 (2000).
  %doi:10.1103/PhysRevC.61.041601
  %%CITATION = doi:10.1103/PhysRevC.61.041601;%%
  %10 citations counted in INSPIRE as of 29 May 2018
  
  %\cite{Wang:2013gk}
\bibitem{Wang:2013gk} 
  S.~M.~Wang, J.~C.~Pei and F.~R.~Xu,
  ``Spectroscopic calculations of cluster nuclei above double shell closures with a new local potential,''
  Phys.\ Rev.\ C {\bf 87}, 014311 (2013).
  %doi:10.1103/PhysRevC.87.014311
  %%CITATION = doi:10.1103/PhysRevC.87.014311;%%
  %4 citations counted in INSPIRE as of 29 May 2018
  
  %\cite{Qian:2018bfu}
\bibitem{Qian:2018bfu} 
  Y.~Qian and Z.~Ren,
  ``Landscape of $\alpha$ preformation probability for even¨Ceven nuclei in medium mass region,''
  J.\ Phys.\ G {\bf 45}, 035103 (2018).
  %doi:10.1088/1361-6471/aaa90b
  %%CITATION = doi:10.1088/1361-6471/aaa90b;%%

%\cite{Ropke:2014wsa}
\bibitem{Ropke:2014wsa} 
  G.~R\"opke {\it et al.},
  ``Nuclear clusters bound to doubly magic nuclei: The case of $^{212}$Po,''
  Phys.\ Rev.\ C {\bf 90}, 034304 (2014)
  %doi:10.1103/PhysRevC.90.034304
  [arXiv:1407.0510 [nucl-th]].
  %%CITATION = doi:10.1103/PhysRevC.90.034304;%%
  %34 citations counted in INSPIRE as of 29 May 2018

%\cite{Xu:2015pvv}
\bibitem{Xu:2015pvv} 
  C.~Xu {\it et al.},
  ``$\alpha$-decay width of $^{212}$Po from a quartetting wave function approach,''
  Phys.\ Rev.\ C {\bf 93}, 011306 (2016)
  %doi:10.1103/PhysRevC.93.011306
  [arXiv:1511.07584 [nucl-th]].
  %%CITATION = doi:10.1103/PhysRevC.93.011306;%%
  %26 citations counted in INSPIRE as of 29 May 2018
  
  %\cite{Xu:2017vyt}
\bibitem{Xu:2017vyt} 
  C.~Xu {\it et al.},
  ``$\alpha$-cluster formation and decay in the quartetting wave function approach,''
  Phys.\ Rev.\ C {\bf 95}, 061306 (2017)
  %doi:10.1103/PhysRevC.95.061306
  [arXiv:1705.00391 [nucl-th]].
  %%CITATION = doi:10.1103/PhysRevC.95.061306;%%
  %11 citations counted in INSPIRE as of 29 May 2018
  
  %\cite{Ropke:2017qck}
\bibitem{Ropke:2017qck} 
  G.~R\"opke {\it et al.},
  ``Alpha-Like Clustering in ${}^{20}$ Ne from a Quartetting Wave Function Approach,''
  J.\ Low.\ Temp.\ Phys.\  {\bf 189}, 383 (2017)
  %doi:10.1007/s10909-017-1796-9
  [arXiv:1707.04517 [nucl-th]].
  %%CITATION = doi:10.1007/s10909-017-1796-9;%%
  %1 citations counted in INSPIRE as of 19 Aug 2018

  
  %\cite{Varga:1992zz}
\bibitem{Varga:1992zz} 
  K.~Varga, R.~G.~Lovas and R.~J.~Liotta,
  ``Absolute alpha decay width of $^{212}$Po in a combined shell and cluster model,''
  Phys.\ Rev.\ Lett.\  {\bf 69}, 37 (1992).
  %doi:10.1103/PhysRevLett.69.37
  %%CITATION = doi:10.1103/PhysRevLett.69.37;%%
  %132 citations counted in INSPIRE as of 29 May 2018
  
  %\cite{Buck:1996zza}
\bibitem{Buck:1996zza} 
  B.~Buck, J.~C.~Johnston, A.~C.~Merchant and S.~M.~Perez,
  ``Cluster model of alpha decay and $^{212}$Po,''
  Phys.\ Rev.\ C {\bf 53}, 2841 (1996).
 % doi:10.1103/PhysRevC.53.2841
  %%CITATION = doi:10.1103/PhysRevC.53.2841;%%
  %48 citations counted in INSPIRE as of 29 May 2018
  
  %\cite{Astier:2009bs}
\bibitem{Astier:2009bs} 
  A.~Astier, P.~Petkov, M.-G.~Porquet, D.~S.~Delion and P.~Schuck,
  ``A novel manifestation of $\alpha$-clustering: new `$\alpha$ + $^{208}$Pb' states in $^{212}$Po revealed by their enhanced E1 decays,''
  Phys.\ Rev.\ Lett.\  {\bf 104}, 042701 (2010)
  %doi:10.1103/PhysRevLett.104.042701
  [arXiv:0911.3502 [nucl-ex]].
  %%CITATION = doi:10.1103/PhysRevLett.104.042701;%%
  %43 citations counted in INSPIRE as of 29 May 2018

%\cite{Ibrahim:2010zz}
\bibitem{Ibrahim:2010zz} 
  T.~T.~Ibrahim, S.~M.~Perez and S.~M.~Wyngaardt,
  ``Hybrid potential model of the alpha-cluster structure of $^{212}$Po,''
  Phys.\ Rev.\ C {\bf 82}, 034302 (2010).
  %doi:10.1103/PhysRevC.82.034302
  %%CITATION = doi:10.1103/PhysRevC.82.034302;%%
  %8 citations counted in INSPIRE as of 29 May 2018
  
  %\cite{Tarbert:2013jze}
\bibitem{Tarbert:2013jze} 
  C.~M.~Tarbert {\it et al.},
  ``Neutron skin of $^{208}$Pb from Coherent Pion Photoproduction,''
  Phys.\ Rev.\ Lett.\  {\bf 112}, 242502 (2014)
  %doi:10.1103/PhysRevLett.112.242502
  [arXiv:1311.0168 [nucl-ex]].
  %%CITATION = doi:10.1103/PhysRevLett.112.242502;%%
  %79 citations counted in INSPIRE as of 29 May 2018

%\cite{Angeli:2013epw}
\bibitem{Angeli:2013epw} 
  I.~Angeli and K.~P.~Marinova,
  ``Table of experimental nuclear ground state charge radii: An update,''
  Atom.\ Data Nucl.\ Data Tabl.\  {\bf 99}, 69 (2013).
  %doi:10.1016/j.adt.2011.12.006
  %%CITATION = doi:10.1016/j.adt.2011.12.006;%%
  %280 citations counted in INSPIRE as of 29 May 2018
  
  \bibitem{Brown:2005}
  B.~A.~Brown, 
  \emph{Lecture Notes in Nuclear Structure Physics}
  (2005).
  
  %\cite{Buck:1975zz}
\bibitem{Buck:1975zz} 
  B.~Buck, C.~B.~Dover and J.~P.~Vary,
  ``Simple potential model for cluster states in light nuclei,''
  Phys.\ Rev.\ C {\bf 11}, 1803 (1975).
  %doi:10.1103/PhysRevC.11.1803
  %%CITATION = doi:10.1103/PhysRevC.11.1803;%%
  %158 citations counted in INSPIRE as of 29 May 2018
  
  %\cite{Michel:1988zz}
\bibitem{Michel:1988zz} 
  F.~Michel, G.~Reidemeister and S.~Ohkubo,
  ``Potential description of the positive- and negative-energy properties of the $\alpha$+ $^{40}$Ca system and $\alpha$-cluster structure of $^{44}$Ti,''
  Phys.\ Rev.\ C {\bf 37}, 292 (1988).
  %doi:10.1103/PhysRevC.37.292
  %%CITATION = doi:10.1103/PhysRevC.37.292;%%
  %37 citations counted in INSPIRE as of 29 May 2018
  
  %\cite{Buck:1995zz}
\bibitem{Buck:1995zz} 
  B.~Buck, J.~C.~Johnston, A.~C.~Merchant and S.~M.~Perez,
  ``Unified treatment of scattering and cluster structure in $\alpha$+closed shell nuclei: $^{20}$Ne and $^{44}$Ti,''
  Phys.\ Rev.\ C {\bf 52}, 1840 (1995).
  %doi:10.1103/PhysRevC.52.1840
  %%CITATION = doi:10.1103/PhysRevC.52.1840;%%
  %28 citations counted in INSPIRE as of 29 May 2018
  
  \bibitem{Ren:1987}
  Z. Ren and G. Xu, 
  ``Reduced alpha transfer rates in a schematic model,''
  Phys.\ Rev.\ C {\bf 36}, 456 (1987).
  
    \bibitem{Ren:1988}
  Z. Ren and G. Xu, 
  ``Evidence of $\alpha$ correlation from binding energies in medium and heavy nuclei,''
  Phys.\ Rev.\ C {\bf 38}, 1078 (1988).











\end{thebibliography}
\end{document}